\documentclass[%
 reprint,
 amsmath,amssymb,
 aps,
]{revtex4-2}

\usepackage{siunitx}
\usepackage{float}
\usepackage{graphicx}
\usepackage{dcolumn}
\usepackage{bm}
\usepackage{graphicx}
\usepackage{color}
\usepackage{dcolumn}
\usepackage{epsfig}
\usepackage{latexsym}
\usepackage{amsfonts}
\usepackage{amssymb}

\usepackage{bbm}
\usepackage{bbold}
\usepackage{mathtools}
\usepackage{booktabs}
\usepackage{placeins} 
\usepackage{stmaryrd} 

\newcommand{\am}{\mathcal{A}} 
\newcommand{\eqdef}{\equiv}
\newcommand{\gaus}{\mathcal{G}} 
\newcommand{\hs}{h_{\rm  s}}
\newcommand{\itf}{-}
\newcommand{\tauh}{\tau_{1/2}}
\newcommand{\taupul}{\tau_\mathrm{pul}}
\newcommand{\tpul}     {t_\mathrm{pul}}
\newcommand{\Tab}{\hat{T}}
\newcommand{\Tin}{T^\mathrm{in}}
\newcommand{\To}{T_\mathrm{o}}

\newcommand{\el}{{\rm  e}}
\newcommand{\ph}{{\rm  p}}
\newcommand{\phs}{{\rm s}}

\newcommand{\bc} {G_{\rm pp}}
\newcommand{\bce}{G_{\rm ep}}
\newcommand{\bcexp}{G^\mathrm{\rm exp}}
\newcommand{\bcth}{G^\mathrm{\rm th}}

\newcommand{\ce}{C_{\el}}
\newcommand{\Cp}{C_{\ph}}
\newcommand{\cs}{C_{\phs}}
\newcommand{\ke}{k_{\el}}
\newcommand{\kp}{k_{\ph}}
\newcommand{\ks}{k_{\phs}}

\newcommand{\kebar}{\bar{k}_\mathrm{e}}
\newcommand{\kpbar}{\bar{k}_\mathrm{p}}

\newcommand{\Gpp}{\bc}
\newcommand{\Gep}{\bce}
\newcommand{\Geff}{G_\mathrm{eff}}
\newcommand{\lep}{\l_\mathrm{ep}}

\newcommand{\etal}{et {\em al}}
\newcommand{\ie}{i.e.}

\newcommand{\Te}{T_{\el}}
\newcommand{\Teab}{\Tab_{\el}}
\newcommand{\Teav}{\overline{T}_{\el}}
\newcommand{\Tein}{\Tin_{\el}}
\newcommand{\Tp}{T_{\ph}}
\newcommand{\Tpav}{\overline{T}_{\ph}}
\newcommand{\Tpin}{\Tin_{\ph}}
\newcommand{\Tpavmax}{\Tpav^\mathrm{max}}
\newcommand{\Ts}{T_{\phs}}
\newcommand{\Tsin}{\Tin_{\phs}}

\newcommand{\bcunit}{MW/m$^2$/K}
\usepackage{color}
\usepackage[dvipsnames]{xcolor}

\newcommand{\corf} [1]{{\color{Black}#1}}

\def\nbOne{{\mathchoice {\rm 1\mskip-4mu l} {\rm 1\mskip-4mu l}                                        {\rm 1\mskip-4mu l} {\rm 1\mskip-4mu l}}}

\begin{document}

\preprint{APS/123-QED}
\title{First-principles calculations of thermal transport at metal/silicon interfaces:\\
evidence of interfacial electron-phonon coupling
}

\author{Michael De San Féliciano$^{1}$}%
\author{Christophe Adessi$^{1}$}%
\author{Julien El Hajj$^{1}$}%
\author{Nicolas Horny$^{2}$}%
\author{François Detcheverry$^{1}$}%
\author{Manuel Cobian$^{3}$}%
\author{Samy Merabia$^{1}$}%
\affiliation{%
 $^{1}$Institut Lumière Matière, UMR5306 Université Claude Bernard Lyon 1-CNRS, Villeurbanne, France \\ 
$^{2}$ ITheMM laboratory, Université de Reims Champagne-Ardennes URCA, Moulin de la Housse, Reims, France \\
 $^{3}$Laboratoire de Tribologie et Dynamique des Systèmes, Ecole Centrale de Lyon-CNRS, Ecully, France}

\date{\today}

\begin{abstract}
With the increasing miniaturization of electronic components and the need to optimize thermal management, it has become essential to understand heat transport at metal-semiconductor interfaces. While it has been recognized decades ago that an electron-phonon channel may take place at metal-semiconductor interfaces, its existence is still controversial. Here, we investigate thermal transport at metal-silicon interfaces using the combination of first-principles calculations and non-equilibrium Green's function (NEGF). 
We explain how to correct NEGF formalism to account for the out-of-equilibrium nature of the energy carriers in the vicinity of the interface. The relative corrections to the equilibrium distribution are shown to arise from the spectral mean free paths of silicon and may reach $15 \%$. Applying these corrections, we compare the predictions of NEGF to available experimental data for Au-Si, Pt-Si and Al-Si interfaces. Based on this comparison, we infer the value of the electron-phonon interfacial thermal conductance by employing the two-temperature model.  
We find that interfacial thermal transport at Au-Si interfaces is mainly driven by phonon-phonon processes, and that electron-phonon processes play a negligible role in this case. 
By contrast, for Al-Si interfaces, we show that phonon-phonon scattering alone can not explain the experimental values reported so far, and we estimate that the electron-phonon interfacial conductance accounts for one third of the total conductance.  
This work demonstrates the importance of the electron-phonon conductance at metal-silicon interfaces and
calls for systematic experimental investigation of thermal transport at these interfaces at low temperatures. It paves the way for an accurate model to predict the conductance associated to the interfacial electron-phonon channel.

\end{abstract}

\maketitle


\section{\label{sec:intro}Introduction}

The miniaturization of electronic devices has continued over the past decades until reaching nanometer sizes, resulting in an increasingly high density of interfaces. Cooling is absolutely necessary to ensure the proper functioning of these devices, 
which requires fast removal of hot spots and optimizing heat evacuation in microprocessors~\cite{management1}. In these systems, heat dissipation in the bulk materials is negligible, and there is a critical need to understand the basic mechanisms of heat transport at interfaces~\cite{management2}.

The central quantity which controls interfacial heat transfer between two media is the interface thermal conductance $G$, also called Kapitza conductance from the name of the Soviet physicist who laid the foundations of interfacial heat transport~\cite{kapitza}. For an interface crossed by a heat flux $\mathcal{J}$ and featuring a temperature jump~$\Delta T$, the interface thermal conductance (ITC) is defined by $G=\mathcal{J}/\Delta T$. Apart from microelectronics and high-power electronic devices, knowledge of the ITC has important consequences for phase-change materials~\cite{wong2010}, wide band gap semiconductor interfaces~\cite{cheng2022},
thermoelectricity~\cite{cahill2014}, batteries~\cite{changqing2022} and nanoparticle-based cancer therapies~\cite{merabia2019}. 



Alongside the development of experimental techniques probing interfacial thermal transport, including the $3\omega$ method \cite{3w} and time-domain thermoreflectance \cite{tdtr}, simulation models have also gained interest for interpreting experimental results. 
Indeed, a full characterization of local thermal transport mechanism is difficult to achieve experimentally, while computational models allow one to investigate the microscopic details of interfacial thermal transport, such as the spectral dependence of phonon transmission.

On the theoretical side, several approaches have been developed to predict the ITC of solid-solid interfaces.
A first level of approximations is found in two semi-empirical models, namely the Acoustic Mismatch Model (AMM)\cite{amm} and the Diffusive Mismatch Model (DMM) \cite{dmm}. These two models are based on a Landauer approach that considers the transmission of phonons emitted by ideal thermostats but ignore the effects of local atomic structure and bonding strength in interfacial thermal transport. A more sophisticated level of description is offered by atomistic simulations, including molecular dynamics (MD) simulations~\cite{landry2009,merabia2012,yang2015,fan2023,elhajj2024} and Green's functions method (NEGF) \cite{negf_intro,agf,wang2008}, which account for the atomistic features of the interface and can predict modal phonon transmission useful to study various types of interfaces. However, MD simulations rely on empirical potentials which may be inadequate in describing optical phonon modes and phonon transmission at interfaces. First-principles calculations based on density functional theory may overcome these limitations at the expense of a heavier computational cost.

One limitation of NEGF is that it treats the contacts as ideal thermostats whose properties are described by equilibrium statistics. However, it has been recently recognized that the non-equilibrium nature of phonons near interfaces may play a significant role~\cite{neq_phonon0}. Indeed, in the vicinity of an interface, the reflection of phonons caused by phonon-interface scattering and the discontinuity of temperature across the interface lead to an out-of-equilibrium state. With the aim of understanding the non-equilibrium behavior of phonons at interfaces, several studies have been carried to infer the modal dependence of the phonon temperature at the interface, including Raman spectroscopy~\cite{neq_exp_phonon1, neq_exp_phonon2} and time-domain thermoreflectance \cite{neq_exp_phonon3,neq_exp_phonon4} experiments, but also computational techniques~\cite{neq_phonon1, neq_phonon2,neq_phonon3,neq_phonon4,neq_phonon5}.

The case of metal-semiconductor interfaces is to be considered separately, as different energy channels may compete in interfacial heat transfer. Along with the phonon-phonon channel, electrons may also couple with the phonons in the semiconductors thus contributing to the total ITC. The existence of this channel has been recognized in the first experimental determination of the Kapitza resistance between two solids at room temperature by Stoner and Marris~\cite{stoner1993}. This seminal work motivated several theoretical developments aimed at predicting the effects of a direct electron-phonon coupling through the interface~\cite{huberman1994,sergeev1998,mahan2009,zhangh2013}. Note that some authors have claimed that the coupling between metal electrons and the semiconductor is indirect resulting in a resistive effect~\cite{majumdar2004,ausi_ref,wang2012b,singh2013}. After the work of Stoner and Marris, a few experimental studies have challenged the existence of a direct coupling~\cite{lyeo2006,hohensee2015} either by investigating thermal transport at semi-metal-semiconductor interface (Bi-Si)~\cite{lyeo2006} or by probing interfacial thermal  transfer under high pressures~\cite{hohensee2015}. However, as argued in Ref.~\cite{lombard2014}, the experimental data on Bi-Si do not preclude any direct energy transfer. 

There are in fact several hints that the direct electron-phonon channel may be non-negligible. First, transient thermoreflectance experiments on a gold film~\cite{hopkins2009}
concluded on the existence of an electron-substrate energy transfer, under strong non-equilibrium conditions~\cite{hopkins2008}. Second, first-principle calculations on Si-CoSi$_2$ interfaces showed evidence of interfacial electron-phonon coupling~\cite{sadasivam2017}. The existence of a coupling should be highly dependent on the precise system under scrutiny, and considerations on some specific systems may not allow one to draw general conclusions. In this context, first-principles calculations may help systematically analyze and quantify the effects of the different energy channels taking place at the interface for a wide class of systems. To this date, however, there have been only few ab-initio studies probably due to the high computational cost of first-principle calculations of interface systems.    

In this work, we investigate thermal transport at three different metal-Si interfaces using a combination of First-principles calculations and non-equilibrium Green's functions. We select three metals, namely Au, Pt and Al, 
whose acoustic contrasts versus silicon are different.
We show that harmonic phonon transport describes well heat transfer measured experimentally at Au-silicon interfaces, suggesting a negligible effect of direct electron-phonon processes in this case. By contrast, we show that harmonic phonon transport is not sufficient to describe heat transfer measured experimentally at Au-Si interfaces, even at low temperatures, and therefore for this system, we conclude that a direct interfacial electron-phonon coupling 
at the interface accounts for one third of the interfacial thermal conductance.

The remainder of this work is structured as follows.
In Sec.~\ref{sec:method}, we present the NEGF method employed to predict thermal transport at metal/silicon interfaces and derive the out-of-equilibrium correction. In Sec.~\ref{sec:results}, we present the results and compare them with available experimental data. The relative importance of interfacial electron-phonon processes is also discussed. Finally, we conclude and give some perspectives in Sec.~\ref{sec:conclusion}.


\section{\label{sec:method}Method}

\subsection{\label{sec:negf} Non-equilibrium Green's functions (NEGF)}

\begin{figure}[h]
  \centering
  \includegraphics[width = 1.0\linewidth]{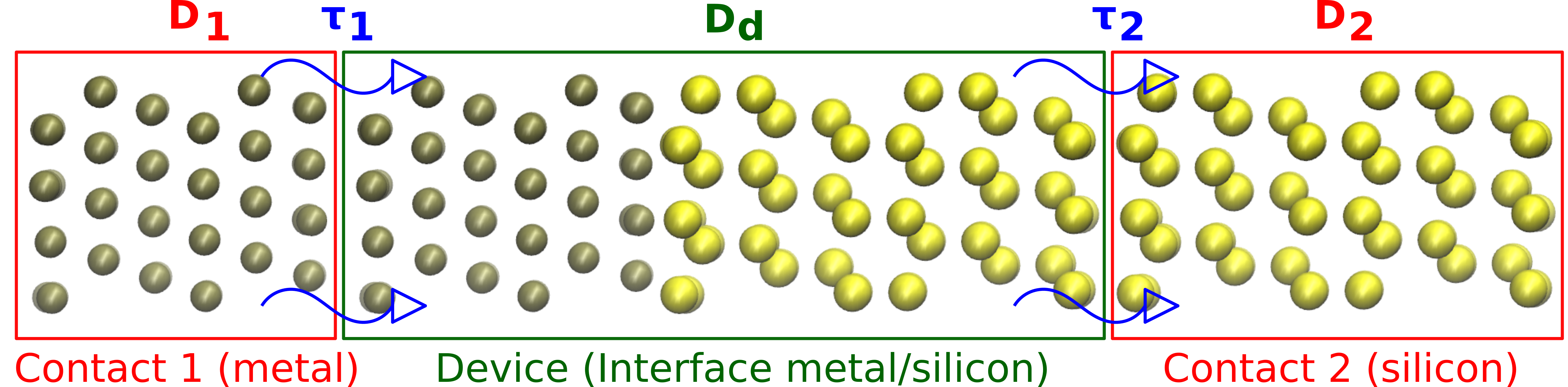}
  \caption{Schematic representation of the metal/semiconductor interface. The system is divided in two contacts and a device. $D_1$, $D_2$ and $D_\mathrm{d}$ represent the dynamic matrices of the isolated contacts and device. $\tau_1$ and $\tau_2$ are the coupling matrices between the contacts and the device.}\label{fig:negf_schema}
\end{figure}

We first review the NEGF approach. Non-Equilibrium  
Green's function technique \cite{agf} is an atomistic method well suited to study phonon or electron transport across an interface. The method assumes that phonons are treated with infinitesimal temperature gradients. Furthermore, this method is mainly employed for the treatment of the harmonic component of interfacial phonon transport and anharmonic phonon scattering is neglected. NEGF allows one to compute the transmission $\mathcal{T}(\omega)$ which subsequently can be employed to obtain the interfacial thermal conductance $G_{\rm eq}$:
\begin{equation}\label{eq:g_green}
  G_{\rm eq} = \int_0^{\infty}\frac{\hbar \omega}{2\pi}\mathcal{T}(\omega)\frac{\partial f_{\rm BE}}{\partial T}d\omega, 
\end{equation}	
where $f_{\rm BE}(\omega,T)$ denotes the Bose-Einstein distribution.
The transmission $\mathcal{T}(\omega)$ is obtained in several steps. The first step relies on the determination of the force constants $\frac{\partial^2 U_{ab}}{\partial u_a \partial u_b}$ obtained from ab-initio calculations where $U_{ab}$ is the interatomic potential between atoms $a$ and $b$ and $u_a$, $u_b$ are the individual spatial displacements of atom $a$ and $b$. The dynamic matrix $D_{ab}$ is built from the force constants as:
\begin{equation}\label{eq:harmo}
{D_{ab}} = \frac{1}{\sqrt{M_a M_b}}
\begin{cases}
\frac{\partial^2 U}{\partial u_a \partial u_b} &\text{if } a\ne b,\\
-\sum\limits_{k\ne a}\frac{\partial^2 U}{\partial u_a \partial u_k} &\text{if } a = b,\\
\end{cases}
\end{equation}
where $M_a$ and $M_b$ are the atomic mass of atoms $a$ and $b$. In the Landauer formalism~\cite{landauer1,landauer2}, the system is divided in two semi-infinite contacts -- a metal bulk and a silicon bulk -- 
and a central region, named the device in the following (Fig.~\ref{fig:negf_schema}).  
The dynamic matrix can be rewritten as:
\begin{gather}
 D_{\rm tot} = 
 \begin{bmatrix} 
 D_1 & \tau_1^{\dagger} & 0 \\ 
 \tau_1 & D_\mathrm{d} & \tau_2 \\
 0 & \tau_2^{\dagger} & D_2
 \end{bmatrix} ,
\end{gather}
where $D_1$, $D_2$ and $D_{\rm d}$ represent the dynamic matrices of the isolated contacts and device (see Fig.~\ref{fig:negf_schema}). $\tau_1$ and $\tau_2$ are the coupling matrices derived from the forces constants between the atoms of the device and those of the contacts. The phonon transmission $\mathcal{T}(\omega)$ can be written as:
\begin{equation}\label{eq:transmission}
  \mathcal{T}(\omega) = \rm{Tr}[\Gamma_1 G_{\rm d} \Gamma_2 G_{\rm d}^{\dagger}],
\end{equation}	
where $G_{\rm d}$ is the Green's function of the system and $\Gamma_i$ the escape rate of the medium $i$ (metal for $i=1$, silicon for $i=2$). $G_\mathrm{d}$ is expressed as:
\begin{equation}\label{eq:g_d}
  G_\mathrm{d} = [\omega^2 \nbOne - D_\mathrm{d} - \Sigma_1 - \Sigma_2]^{-1},
\end{equation}	
where $\omega$ is the angular frequency, $\nbOne$ the unit tensor and $\Sigma_i$ the self-energy of medium~$i$. 
The self-energies are obtained from the Green's functions of the contacts alone $G_1$ and $G_2$:
\begin{equation}\label{eq:self_energie}
\Sigma_i = \tau_i G_i \tau_i^{\dagger},
\end{equation}	
and these Green's functions are defined as
\begin{equation}\label{eq:green_contact}
  G_i = [\omega^2 \Gamma_1 - D_i]^{-1}.
\end{equation}	
Finally, the escape rate $\Gamma_i$ can be expressed as:
\begin{equation}\label{eq:rate}
  \Gamma_i = \sqrt{-1}(\Sigma_i-\Sigma_i^{\dagger}).
\end{equation}	
This completes the computation of the transmission as a function of the interatomic force constants.

\subsection{\label{sec:ne_corr}Non-equilibrium corrections}

\subsubsection{General expression of non-equilibrium corrections}
NEGF traditionally assumes ideal contacts in which the excitations (here the phonons) are assumed to be at equilibrium. Here, we derive the non-equilibrium corrections that need to be implemented within the NEGF formalism. It is important to mention that here and in all the following, we concentrate solely on the phonon contribution to the heat flux.  

The heat flux, within the Landauer formalism, is given by:
\begin{equation}\label{eq:j}
  J =\int_0^{\infty} \frac{\hbar \omega}{2\pi} \mathcal{T}(\omega)[N_1(\omega)-N_2(\omega)]d\omega.
\end{equation}  
In this expression, $N_1(\omega)$ and $N_2(\omega)$ are the phonon distributions in contact 1 (metal) and contact~2 (silicon). At equilibrium, $N_1(\omega)=f_{\rm BE}(\omega,T_1)$ and $N_2(\omega)=f_{\rm BE}(\omega,T_2)$, where $f_{\rm BE} (\omega,T)$ denotes the Bose-Einstein distribution of phonons at temperature $T$. Equation~(\ref{eq:j}) does not account, however, for the presence of a temperature gradient. This approximation leads to: 
\begin{equation}\label{eq:j_approx}
  J \simeq \int_0^{\infty} \frac{\hbar \omega}{2\pi} \mathcal{T}(\omega) {\Big(\frac{\partial f_{BE}}{\partial T}\Big)}_{\bar T} d\omega (T_1-T_2) = G_{\rm eq} (T_1-T_2)
\end{equation}
where $\bar T=(T_1+T_2)/2$. To take into account non-equilibrium corrections to the interfacial conductance, we use a first order expansion of $N_1$ and $N_2$ as:
\begin{subequations} \corf{
\begin{align} 
  &N_1(\omega)=f_{\rm BE}(\omega,T_1)+\delta f_1(\omega,T_1),    \\
  &N_2(\omega)=f_{\rm BE}(\omega,T_2)+\delta f_2(\omega,T_2),      
\end{align}
}
\end{subequations}
where $\delta f_1(\omega,T_1)$ and $\delta f_2(\omega,T_2)$ are deviations to the Bose-Einstein distribution. They can be obtained from the steady state solution of the Boltzmann equation $\frac{\partial f_{\bm {q}\nu}}{\partial t}=0$ which implies: 
\begin{equation}
\label{eq:dev_boltz1}
  \left( \frac{ \partial f_{\rm BE}}{\partial T} \right) \, {\bm v}_{{\bm q}\nu}\cdot {\bm \nabla}T \approx -\frac{\delta f_{{\bm q}\nu}}{\tau_{{\bm q}\nu}},
\end{equation}	
where $\tau_{{\bm q}\nu}$ and ${\bm v_{{\bm q}\nu}}$ are respectively the phonon lifetime and the group velocity of the phonon mode characterized by the wave vector ${\bm q}$ and the polarization $\nu$. Assuming diffusive phonon transport in the contact $i$, we have ${\bm J}=-\lambda_i {\bm \nabla}T$ where $\lambda_i$ is the phonon thermal conductivity of the bulk contact $i$. It leads to:
\begin{equation}\label{eq:dev_boltz2}
\delta f_{{\bm q}\nu} = \frac{\tau_{{\bm q}\nu}}{\lambda_i}\Big(\frac{\partial f_{\rm BE}}{\partial T}\Big)  \, {\bm v}_{{\bm q}\nu}\cdot {\bm J}.
\end{equation}
\corf{We can introduce the spectral deviations of the contact $\delta f_i$ which is defined as:}
\begin{subequations}
\corf{
\begin{align}
   \delta f_1(\omega,T)g_1(\omega) &= \sum_{{\bm q}\nu}^{+} \delta f_{{\bm q}\nu} \delta(\omega-\omega_{{\bm q}\nu}), \\
    \delta f_2(\omega,T)g_2(\omega) &= \sum_{{\bm q}\nu}^{-} \delta f_{{\bm q}\nu} \delta(\omega-\omega_{{\bm q}\nu}),
\label{eq:dev_boltz3}
\end{align}
}
\end{subequations}
where $g_i(\omega) = \sum\limits_{q\nu}\delta(\omega-\omega_{q\nu})$ 
is the phonon density of states of contact $i$. 
\corf{The symbols "+" and "-" mean that the sum is restricted to modes propagating toward the interface with $v_z > 0$ or $v_z <0 $ respectively, where $z$ is the direction of the flux $J$. 
By combining Eq.~\eqref{eq:dev_boltz2} and Eq.~\eqref{eq:dev_boltz3}, we get:
\begin{subequations}\label{eq:dev_boltz4}
\begin{align}
    \delta f_1(\omega,T)= &\frac{1}{\lambda_1 g_1(\omega)}\Big(\frac{\partial f_{BE}}{\partial T}\Big)  \nonumber \\
    & \times \Big(\sum_{{\bm q}\nu}^{+}\tau_{{\bm q}\nu}v_{z,{\bm q}\nu}\delta(\omega-\omega_{{\bm q}\nu})\Big)J. \\
     \delta f_2(\omega,T)= &\frac{1}{\lambda_2 g_2(\omega)}\Big(\frac{\partial f_{BE}}{\partial T}\Big)  \nonumber \\
    & \times \Big(\sum_{{\bm q}\nu}^{-}\tau_{{\bm q}\nu}v_{z,{\bm q}\nu}\delta(\omega-\omega_{{\bm q}\nu})\Big)J.
\end{align}
\end{subequations}	
In this latter expression, we can identify the mean free path of the contact $i$ as:
\begin{equation}\label{eq:dev_boltz5}
    \Lambda_i^{\pm}(\omega)=\frac{1}{g_i(\omega)}\sum_{{\bm q}\nu}^{\pm}\tau_{{\bm q}\nu}v_{z,{\bm q}\nu} \delta(\omega-\omega_{{\bm q}\nu}).
\end{equation}
}
If we replace the term of Eq.~\eqref{eq:dev_boltz5} in  Eq.~\eqref{eq:dev_boltz4}, we obtain:
\corf{
\begin{subequations}\label{eq:dev_boltz}
\begin{align}
  \delta f_1(\omega,T) &=\frac{J}{\lambda_1}\Big(\frac{\partial f_{BE}}{\partial T}\Big)\Lambda_1^{+}(\omega),  \\
  \delta f_2(\omega,T) &=\frac{J}{\lambda_2}\Big(\frac{\partial f_{BE}}{\partial T}\Big)\Lambda_2^{-}(\omega) = -\frac{J}{\lambda_2}\Big(\frac{\partial f_{BE}}{\partial T}\Big)\Lambda_2^{+}(\omega), 
  \end{align}
\end{subequations}
where in the second line, we have used the property of isotropy of the bulk leads, which implies $\sum\limits_{{\bm q}\nu}^{-}v_{{\bm q}\nu_z}\tau_{{\bm q}\nu} = -\sum\limits_{{\bm q}\nu}^{+}v_{{\bm q}\nu_z}\tau_{{\bm q} \nu}$ and consequently $\Lambda_i^{-}(\omega)=-\Lambda_i^{+}(\omega)$. In the following, we define for each lead $\Lambda_i(\omega)=\Lambda_i^{+}(\omega)-\Lambda_i^{+}(\omega)$
where~:
\begin{equation}\label{eq:dev_boltz6}
    \Lambda_i(\omega)=\frac{1}{g_i(\omega)}\sum_{{\bm q}\nu}\tau_{{\bm q}\nu}\vert v_{z,{\bm q}\nu} \vert \delta(\omega-\omega_{{\bm q}\nu}).
\end{equation}
so that the out-of-equilibrium distributions become~:
\begin{subequations}\label{eq:dev_boltz}
\begin{align}
  \delta f_1(\omega,T) &=\frac{J}{2\lambda_1}\Big(\frac{\partial f_{BE}}{\partial T}\Big)_{T_1}\Lambda_1(\omega),  \\
  \delta f_2(\omega,T) &=-\frac{J}{2\lambda_2}\Big(\frac{\partial f_{BE}}{\partial T}\Big)_{T_2}\Lambda_2(\omega)  
  \end{align}
\end{subequations}

} 

The corrections for the phonon distributions inserted in Eq.~\eqref{eq:j} lead to the following expression of the heat flux:
\corf{
\begin{equation}\label{eq:flux_correc}
\begin{aligned}
J =& \int_0^{\infty} \frac{\hbar \omega}{2\pi} \mathcal{T}(\omega) \Big[\left(f_{BE}(\omega,T_1)-f_{BE}(\omega,T_2)\right)\\
& + \left(\frac{1}{2\lambda_1}\Big(\frac{\partial f_{BE}}{\partial T}\Big)_{T_1}\Lambda_1(\omega) + \frac{1}{2\lambda_2}\Big(\frac{\partial f_{BE}}{\partial T}\Big)_{T_2}\Lambda_2(\omega)\right) J \Big]d \omega.
\end{aligned}
\end{equation}}
A further approximation is: 
\begin{equation}\label{eq:approx_dfbe}
   \Big(\frac{\partial f_{\rm BE}}{\partial T}\Big)_{T_i} \simeq \Big(\frac{\partial f_{\rm BE}}{\partial T}\Big)_{\bar T} \simeq \frac{f_{\rm BE}(\omega,T_1)-f_{\rm BE}(\omega,T_2)}{T_1 - T_2}.
\end{equation}	
Including this approximation in Eq.~\eqref{eq:approx_dfbe} allows to rewrite the heat flux of Eq.~\eqref{eq:flux_correc} with $G_{\rm eq}$:
\begin{equation}\label{eq:gneq}
\begin{aligned}
  &J = G_{\rm neq}(T_1-T_2) \quad \text{with }G_{\rm neq} = \frac{G_{\rm eq}}{1-\frac{\beta_1}{\lambda_1}-\frac{\beta_2}{\lambda_2}},
\end{aligned}
\end{equation}	
where $\beta_i$ is defined as:
\begin{equation}\label{eq:beta_i}
  \beta_i = \int_0^{\infty} \frac{\hbar\omega}{4\pi}\mathcal{T}(\omega)\Lambda_i(\omega)\Big(\frac{\partial f_{BE}}{\partial T}\Big)d\omega.
\end{equation}	
To go further, we chose to express the phonon thermal conductivity $\lambda_i$ through $\beta_i$. 
Indeed, for the interface between two identical media $i$, we should have no resistance to heat flow, namely $G_{\rm neq} \rightarrow \infty$, meaning:
\begin{equation}\label{eq:g_inf_dev}
\begin{aligned}
  &1-\frac{\beta_i}{\lambda_i}-\frac{\beta_i}{\lambda_i} = 0 \iff \beta_i = \frac{\lambda_i}{2},\\
  &\text{with } \beta_i = \int_0^{\infty} \frac{\hbar \omega}{4\pi} \mathcal{T}_i(\omega) \Lambda_i(\omega)\Big(\frac{\partial f_{BE}}{\partial T}\Big)d\omega,
\end{aligned}
\end{equation}	
where $\mathcal{T}_i(\omega)$ corresponds to the transmission across the fictive interface between the two leads $i$. We can thus write:
\begin{equation}\label{eq:b_l}
  \frac{\beta_i}{\lambda_i} = \frac{\int_0^{\infty} \frac{\hbar\omega}{4\pi}\mathcal{T}(\omega)\Lambda_i(\omega)\Big(\frac{\partial f_{\rm BE}}{\partial T}\Big)d\omega} { \int_0^{\infty} \frac{\hbar\omega}{2\pi}\mathcal{T}_i(\omega)\Lambda_i(\omega)\Big(\frac{\partial f_{\rm BE}}{\partial T}\Big)d\omega}. 
\end{equation}	
To estimate the correction term $\beta_i/\lambda_i$, we need the expression of the spectral phonon lifetimes $\Lambda_i(\omega)$ in the contact $i$. In the following, we will propose three different approaches to compute the spectral phonon lifetimes.
\subsubsection{\label{sec:call}Callaway's model}

The ratio $\beta_i/\lambda_i$ can be simplified using the  popular Callaway's model~\cite{gru_theory} which assumes the spectral mean free paths is described as:
\begin{equation}\label{eq:mfp_cal}
  \Lambda_{{\bm q}\nu} \approx \frac{ v_{{\bm q}\nu} A(T)}{\omega^2},
\end{equation}	
where $A(T)$ is a temperature dependent material constant.
In the limit of low frequency, we express the density of state using Debye's model:
\begin{equation}\label{eq:b_l_cal1}
  g(\omega) = \sum_{{\bm q} \nu} g_{{\bm q}\nu}(\omega) \approx \omega^2 \sum_{{\bm q}\nu} \frac{1}{v_{{\bm q}\nu}^3} = \frac{\omega^2}{3v_{ac}^3},
\end{equation}	
where $v_{ac}$ is the average sound velocity. By combining Eq.~\eqref{eq:mfp_cal} and Eq.~\eqref{eq:b_l_cal1}, we get:
\begin{subequations}\label{eq:b_l_cal2}
\begin{align}
  \Lambda(\omega) &= \frac{1}{g(\omega)}\sum_{{\bm q}\nu} \frac{v_{{\bm q}\nu} A(T)}{\omega^2} g_{q\nu}(\omega), \\
  &= \frac{3 v_{ac}^3}{\omega^2}\sum_{{\bm q}\nu} v_{{\bm q}\nu} A(T).
\end{align}
\end{subequations}	
We thus obtain: 
\begin{equation}\label{eq:b_l_cal}
\begin{aligned}
    \frac{\beta_i}{\lambda_i} &= \frac{\int_0^{\infty} \frac{\hbar \omega}{4\pi} \frac{3v_{ac}^3}{\omega^2}\sum_{q\nu} v_{q\nu} A(T) \mathcal{T}(\omega) k_B d\omega}{\int_0^{\infty} \frac{\hbar \omega}{2\pi} \frac{3v_{ac}^3}{\omega^2}\sum_{q\nu} v_{q\nu} A(T) \mathcal{T}_i(\omega) k_B d\omega},\\[10pt]
    &= \int_0^{\infty}\frac{\mathcal{T}(\omega)}{\omega}d\omega\Big/\int_0^{\infty}\frac{2\mathcal{T}_i(\omega)}{\omega}d\omega.
\end{aligned}
\end{equation}	
Within the Callaway approximation, we only need the knowledge of the two transmissions 
$\mathcal{T}(\omega)$ and $\mathcal{T}_i(\omega)$ to calculate the non-equilibrium correction and no other terms are needed to estimate the correction.

\subsubsection{\label{sec:gru}Grüneisen model}
As an alternative to Callaways's model, the mean free paths may be estimated using the value of the Grüneisen parameter, which for a phonon mode ${\bm q}\nu$, is defined as~\cite{gru_theory}:
\begin{equation}\label{eq:gru}
    \gamma_{{\bm q}\nu}=-\frac{V}{\omega_{{\bm q}\nu}}\left(\frac{\partial \omega_{{\bm q}\nu}}{\partial V}\right)_T,
\end{equation}	
where $V$ is the volume of the system. The Grüneisen parameter is dimensionless and represents the volume dependence of phonon frequencies. Its calculation only requires harmonic phonon calculations. The mean free paths can be approximated using this parameter as~\cite{gru_theory,adessi_gru}: 
\begin{equation}\label{eq:mfp_gru}
    \Lambda_{{\bm q}\nu}(\omega)=v_{{\bm q}\nu}\frac{M}{k_B T}\frac{v^2_{{\bm q}\nu}}{\gamma^2_{{\bm q}\nu}}\frac{\omega^{\rm max}_{\nu}}{\omega^2},
\end{equation}	
where $\omega^{\rm max}_{\nu}$ is the maximal frequency of mode $\nu$.

\subsubsection{\label{sec:pho}Three phonon scattering calculations (3-Ph)}

A third possibility is to perform the explicit calculation of phonon lifetimes $\tau_i(\omega)$ based on three phonon (3-Ph) scattering. The phonon lifetime is related to mean free paths as usual by:
\begin{equation}\label{eq:mfp_tp}
    \Lambda_{{\bm q}\nu} (\omega) = v_{{\bm q}\nu}(\omega)\tau_{{\bm q}\nu}(\omega).
\end{equation}	
This latter quantity may be computed using ab-initio calculations of three-phonon scattering of the bulk materials.

\section{\label{sec:results}Results}

\subsection{\label{sec:dft}DFT calculations}

We investigated three different interfaces: Au/Si, Pt/Si and Al/Si. These three metal systems have been chosen because of they are characterized by different Debye frequencies $\omega_D$~:~$\omega_D= 5$ THz for Au, $\omega_D= 6$ THz for Pt and $\omega_D= 11$ THz for Al and therefore they display different acoustic contrast with silicon. 
Each of these interfaces is composed of two media: a metal (Au, Pt or Al) consisting of a bulk of $4\times4\times3$ primitive cells corresponding to a total of 96 atoms; and a silicon bulk of $3\times3\times3$ primitive cells corresponding to a total of $108$ atoms. The number of cells along the $x$ and $y$ directions (parallel to the interface) was chosen to allow a compromise between a small system (to limit calculation times) and an acceptable matching of the global lattice parameters between metal and silicon bulk along the $x$ and $y$ directions. The number of cells along the thermal transport direction $z$ was chosen to reach a compromise between a relative small system but large enough for NEGF transport calculations. 

These three systems were optimized using SIESTA~\cite{siesta1} under the generalized gradient approximation~\cite{gga} and using Troullier-Martin norm-conserving pseudopotentials~\cite{pseudo}. \corf{The initial structure has been allowed to relax in order to minimize the interatomic forces. During this relaxation step, the maximum force tolerance was $10^{-4}$ eV$/$\si{\angstrom} and the maximum stress tolerance 100 bar.} A Monkhorst-Pack of $3\times3\times1$ \textit{k}-points for Au/Si and Pt/Si and $6\times6\times1$ \textit{k}-points for Al/Si, have been used for the calculations using a mesh cutoff of $400$ Ry. The basis used correspond to a single-zeta plus polarisation basis optimized using the simplex tool of the SIESTA package.

\begin{figure}[h]
  \centering
  \includegraphics[width = 0.7\linewidth]{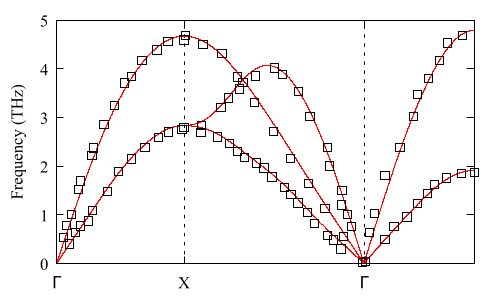}
  \includegraphics[width = 0.7\linewidth]{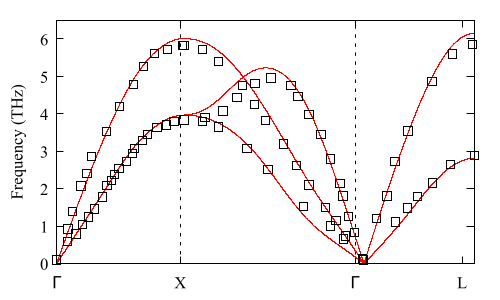}
  \includegraphics[width = 0.7\linewidth]{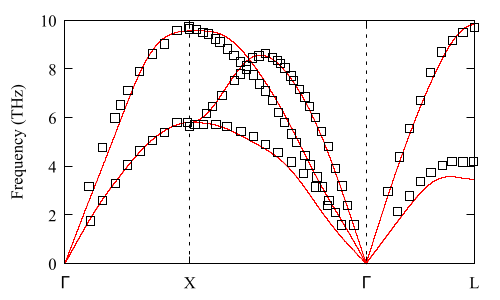}
  \includegraphics[width = 0.7\linewidth]{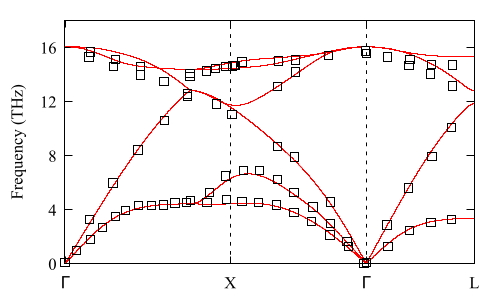}
  \caption{Bulk phonon dispersion curves obtained by DFT calculations (lines) compared to experimental data (symbols). From top to bottom: Au, Pt, Al and Si. The experimental data are respectively taken from Refs.~\cite{disp_Al,disp_Pt,disp_Au,disp_Si_1,disp_Si_2}.}\label{fig:disp_pho}
\end{figure}	

In practice, the force constants were calculated using the phonopy package~\cite{phonopy1,phonopy2} which is based on a supercell approach with the finite displacement method. The supercell used here from the previous optimized one is $3\times3\times1$. The same calculation was done on the bulk of metal ($4\times4\times3$ supercell) and silicon ($3\times3\times3$ supercell) in order to define the leads within the Landauer formalism. The force constants were later used within the Landauer formalism described in Sec.~\ref{sec:method} to compute the phonon transmission.

The bulk phonon dispersion curves of silicon and of the three metals 
are displayed in Fig.~\ref{fig:disp_pho}. Overall, the dispersion curves compare well to experimental data. The only slight discrepancies are for the bulk Al transverse branch close to the L-point, but the corresponding difference between DFT calculations and experimental data is relatively small, typically less than $15 \%$.

\subsection{\label{sec:negf_cal}NEGF calculations}

\begin{figure}[h]
  \centering
  \includegraphics[width = 1.0\linewidth]{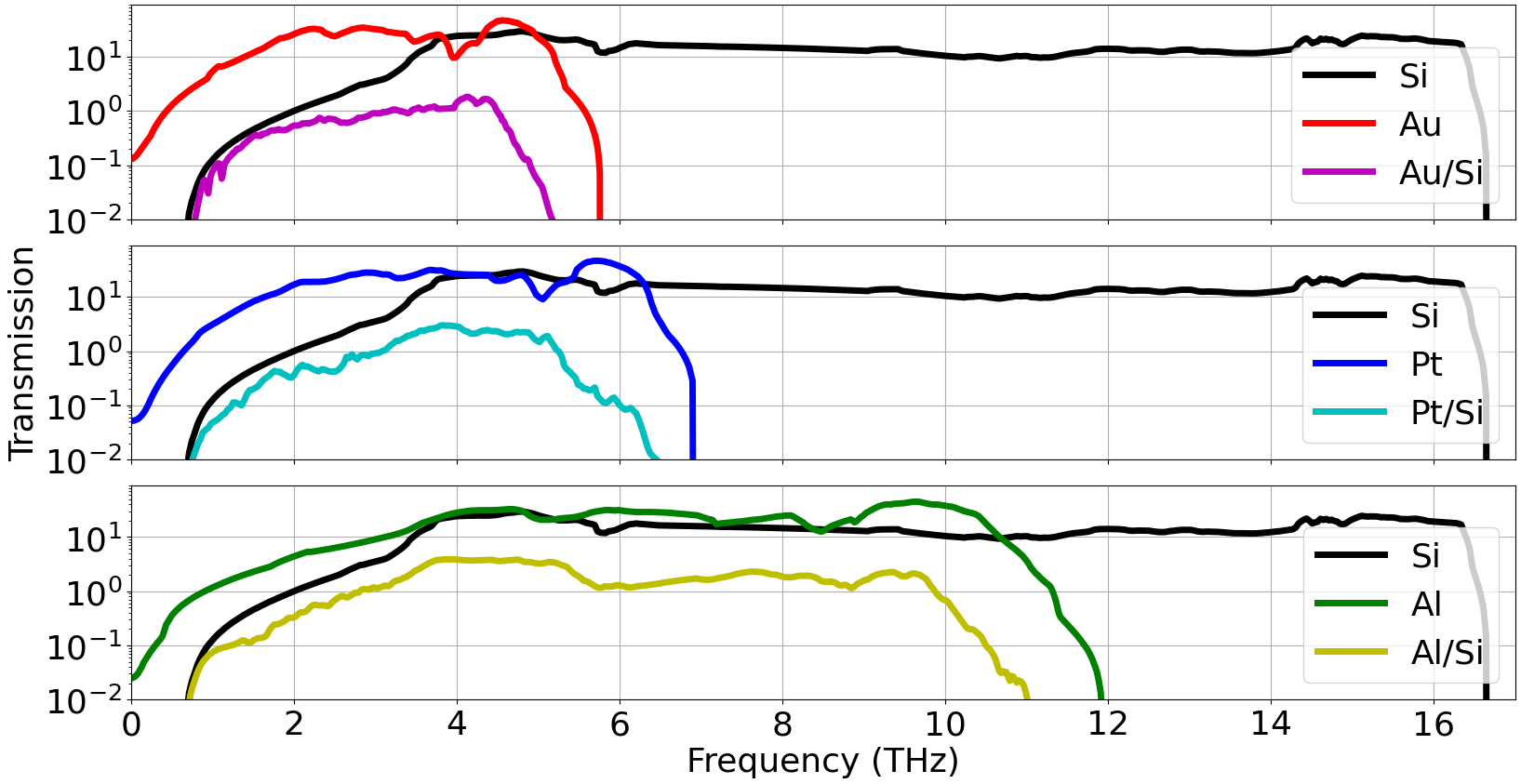}
  \caption{Phonon-phonon transmission calculated within the Landauer formalism for the three systems considered.}\label{fig:trans}
\end{figure}	

The transmission calculated for the different systems using SIESTA and the Landauer formalism is represented in Fig.~\ref{fig:trans}. These transmissions provide information on the phonon frequency ranges involved in phonon interfacial transport. These transmissions allow us to calculate the interfacial phonon-phonon conductance $G_{\rm eq}$, using Eq.~\eqref{eq:g_green}. 

\begin{figure}[h]
  \centering
  \includegraphics[width = 1.0\linewidth]{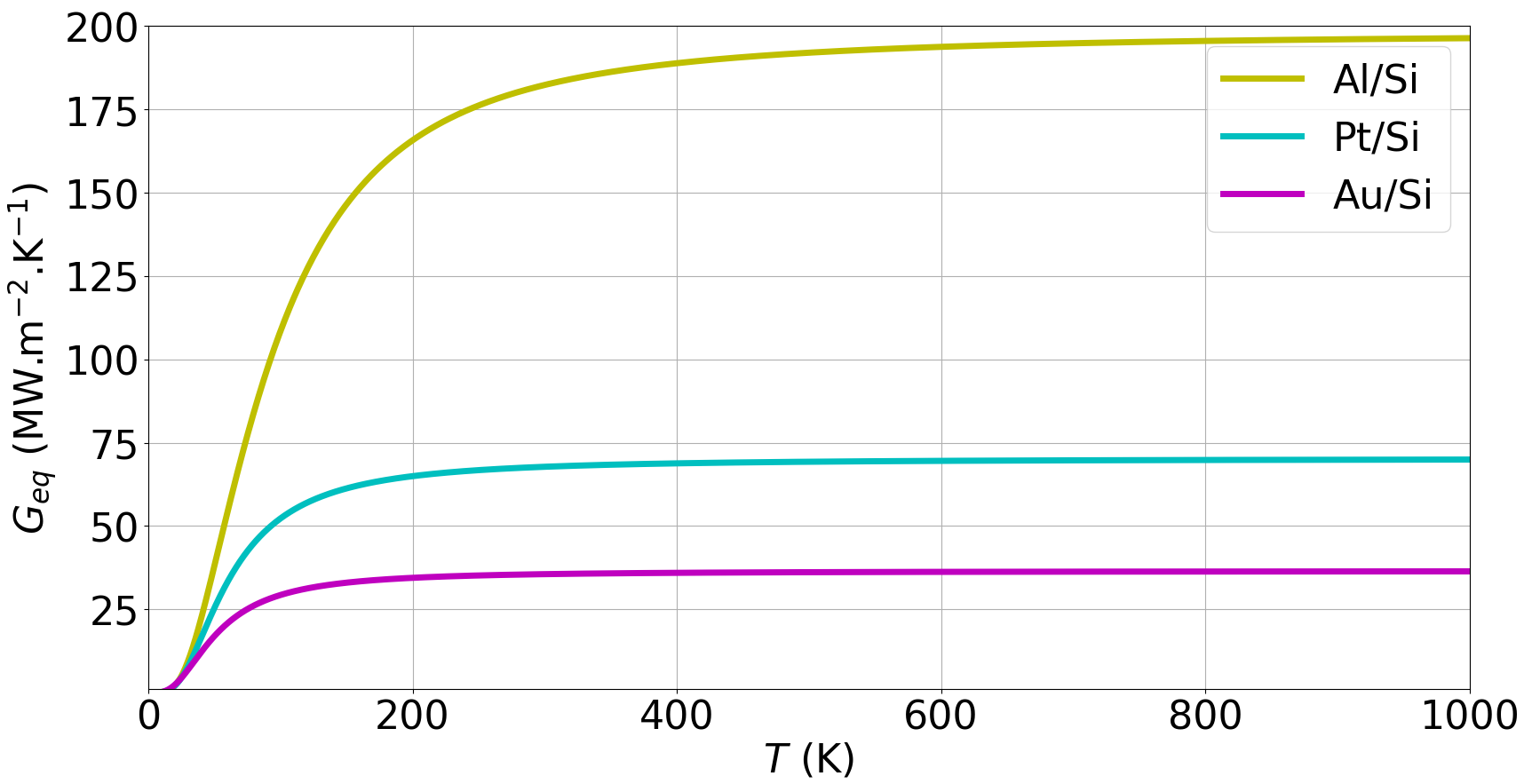}
  \caption{Interfacial phonon-phonon conductance as a function of temperature computed with the NEGF method for the three interfaces Au/Si, Pt/Si and Al/Si.}\label{fig:cond_temp}
\end{figure}	

The frequency range associated with the transmissions at the interfaces is limited by the overlap of the frequency ranges of the metals and silicon. This observation also explains the differences in conductance obtained and represented in Fig.~\ref{fig:cond_temp}. Indeed, the conductance of the Al/Si interface is much higher than the two others and this difference is related to the high phonon cut-off frequency of aluminum resulting in a relative broad width of the frequency range for phonon transmission at the Al/Si interface. Finally, one should note that silicon has a large phonon cut-off frequency $\omega_{\rm cut}^{\rm Si}$, around $16$ THz, and is higher than the cut-off frequencies of the other metals. Moreover, it should be kept in mind that the overlap of phonon modes that effectively transmit the energy is bounded by $\omega_{\rm cut}^{Si}$ is due to the harmonic nature of phonon transport assumed in the present NEGF calculations.

\subsection{\label{sec:ne_cal}Non-equilibrium corrections}

The conductances shown in Fig.~\ref{fig:cond_temp} yield a first insight in the interfacial transport properties as they rely on the assumption of contacts held at equilibrium. The influence of non-equilibrium corrections need to be evaluated. As detailed in Sec.~\ref{sec:method}, there are different routes to estimate the spectral phonon mean free paths. In the case of the Callaway formulation, the corrections are directly deduced from the transmissions presented in Fig.~\ref{fig:trans} and obtained with the NEGF method. Conversely, corrections with the Grüneisen model require additional ab-initio calculations in order to determine the Grüneisen parameter used in the calculation of the mean free path by Eq.~\eqref{eq:mfp_gru}. 

\begin{figure}[h]
  \centering
  \includegraphics[width = 1.0\linewidth]{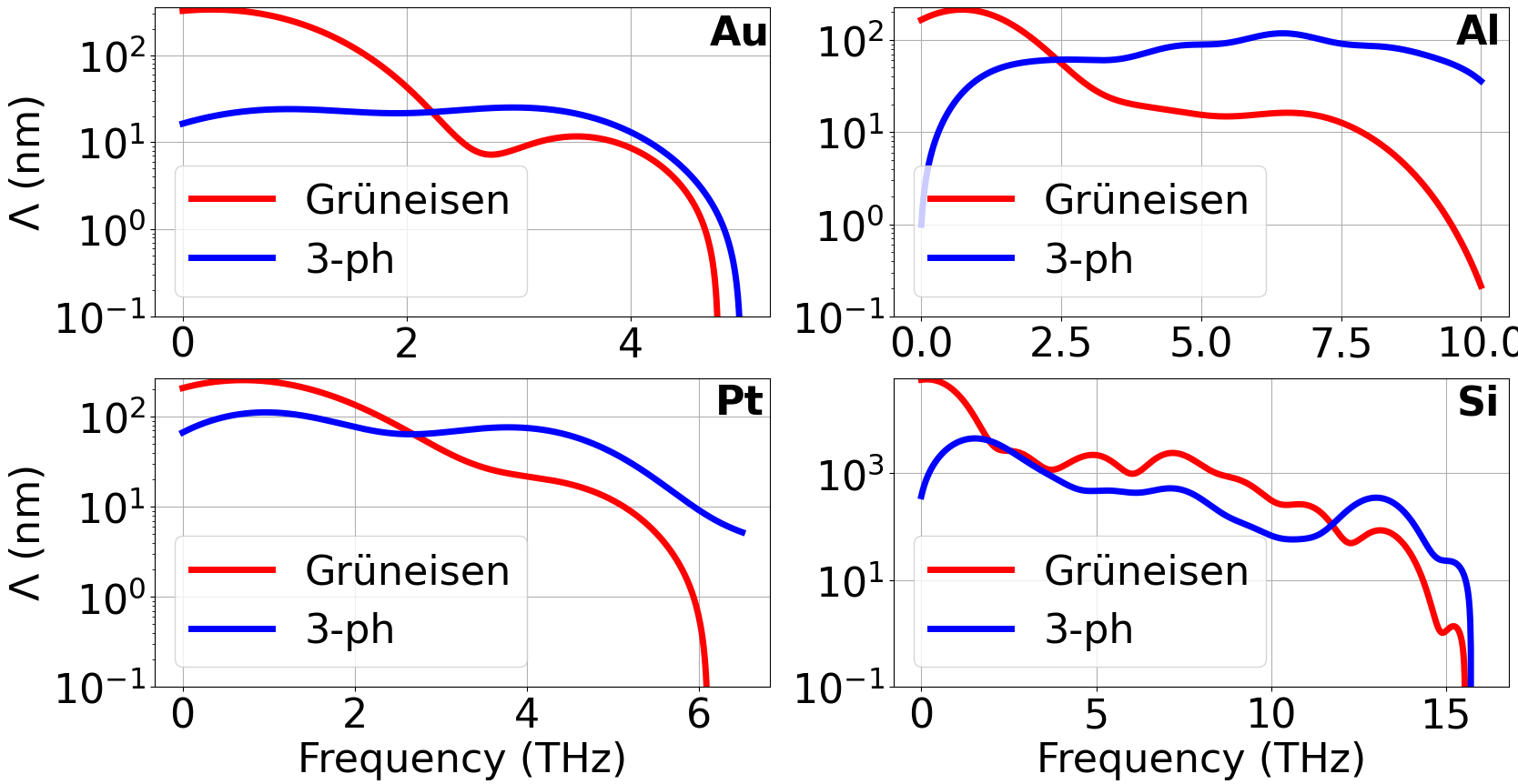}
  \caption{Spectral phonon mean free paths $\Lambda_i$, smoothed out by the gaussian regressor~\cite{gauss_ml}, and calculated using the Grüneisen parameter and the $3$ phonon lifetimes calculations.
   }\label{fig:mfp_ml}
\end{figure}

To estimate these corrections, we have optimised a bulk of one primitive cell for each metal and silicon lead using SIESTA. After optimisation, we created and optimised two other systems for each medium at $99\%$ and $101\%$ of the optimal lattice parameter. The force constants and the Grüneisen parameter for each medium were computed with phonopy. The values of the Grüneisen parameters shown in the Supp. Mat.~\cite{supp_mat} can be compared to the data from Refs.~\cite{gru_Al,gru_Si}. This parameter being linked to the phonon mean free path $\Lambda_i$, the correction of the conductance is carried out by post-processing based on Eq.~\eqref{eq:mfp_gru}.
The resulting spectral mean free paths are shown in Fig.~\ref{fig:mfp_ml}.
 Similarly, the approach based on three phonon scattering requires additional calculations here necessary to determine the phonon lifetimes entering in the calculation of the mean free path through Eq.~\eqref{eq:mfp_tp}. The phonon lifetimes are computed using phono3py~\cite{phono3py}, which allows the calculation of three phonon scattering lifetimes based on the three body force constants characterizing the optimized bulk systems build for the calculation of the Grüneisen parameters. These lifetimes are shown in the Supp. Mat.~\cite{supp_mat} and can be compared to the data from Ref.~\cite{lifetime} for aluminum and silicon. The corresponding spectral mean free paths are displayed in Fig.~\ref{fig:mfp_ml}.

\begin{figure}[h]
  \centering
  \includegraphics[width = 1.0\linewidth]{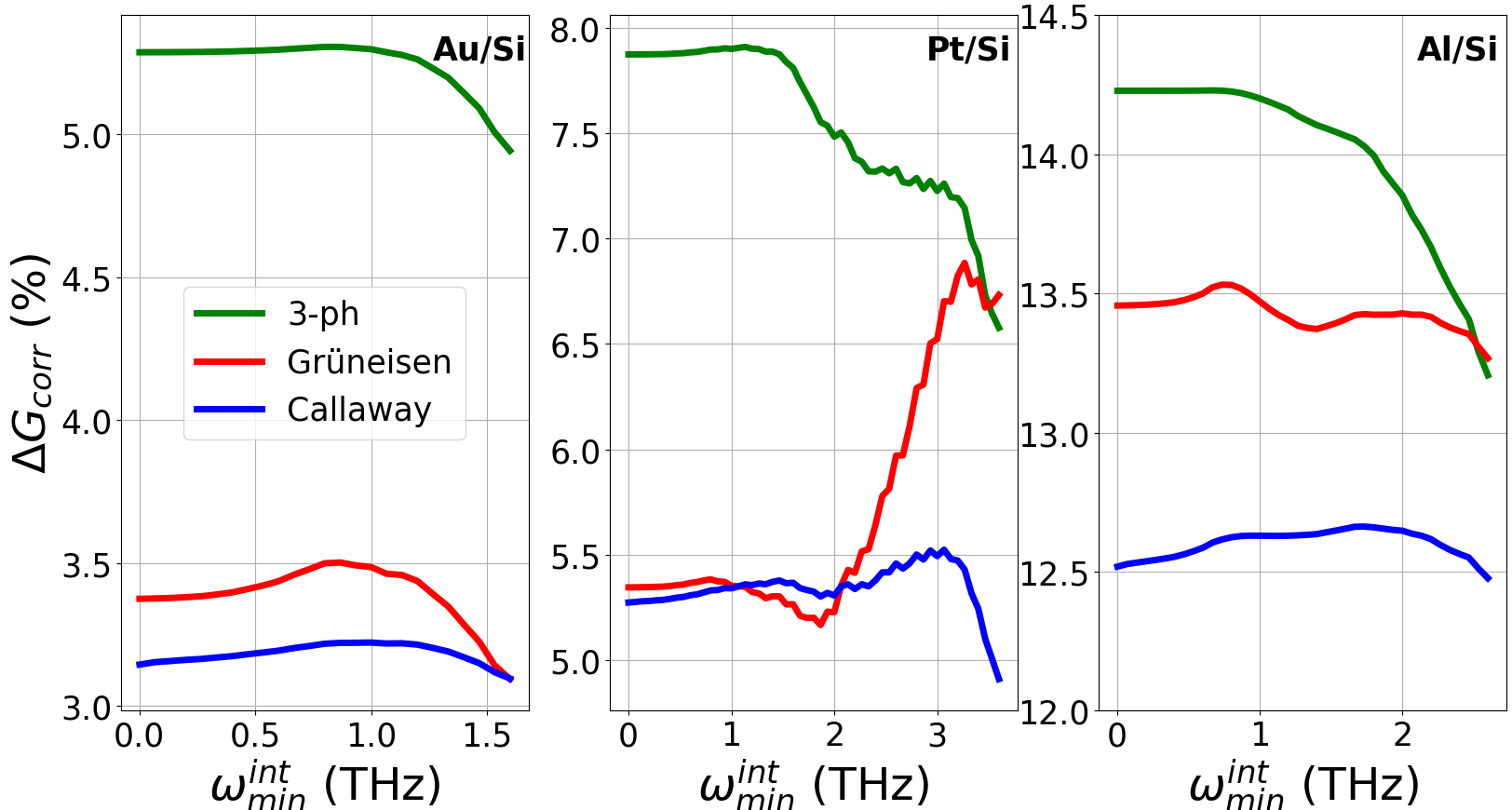}
  \caption{Relative deviation of the conductance $\Delta G_{\rm corr} = \Big(G_{\rm neq}/G_{\rm eq}-1\Big)$ due to non-equilibrium corrections at $300\,$K. The non-equilibrium deviations are calculated using Callaway's model, Grüneisen parameter, and three phonon scattering (3-Ph) calculations. The abscissa $\omega_{\rm min}^{\rm int}$ is relative to the lower bound of the integrals of Eq.~\eqref{eq:b_l}.}\label{fig:ne_corr}
\end{figure}	

From these three non-equilibrium corrections, we determined a correction associated with each of the three interfaces Au/Si, Pt/Si and Al/Si. The results are presented in Fig.~\ref{fig:ne_corr}. First, we can see that the corrections can be relatively important depending on the interface, around $15 \%$ for Al/Si and therefore should not be neglected. The corrections are presented as a function of $\omega_{\rm min}^{\rm int}$ which refers to the lower bound of the frequencies used in the calculations of the integrals in Eq.~\eqref{eq:b_l}. This term is a metric to ensure that we are not missing important low-frequency modes. 

By plotting the correction as a function of $\omega_{\rm min}^{\rm int}$, we can estimate the weight represented by this low density of points at low frequency. The value of the non-equilibrium correction does not vary significantly when $\omega_{\rm min}^{\rm int}$ is below $0.5\,$THz. This means that cutting phonon modes with low frequency does not affect too much the calculation of the non-equilibrium corrections. 

\begin{figure}[h]
  \centering
  \includegraphics[width = 1.0\linewidth]{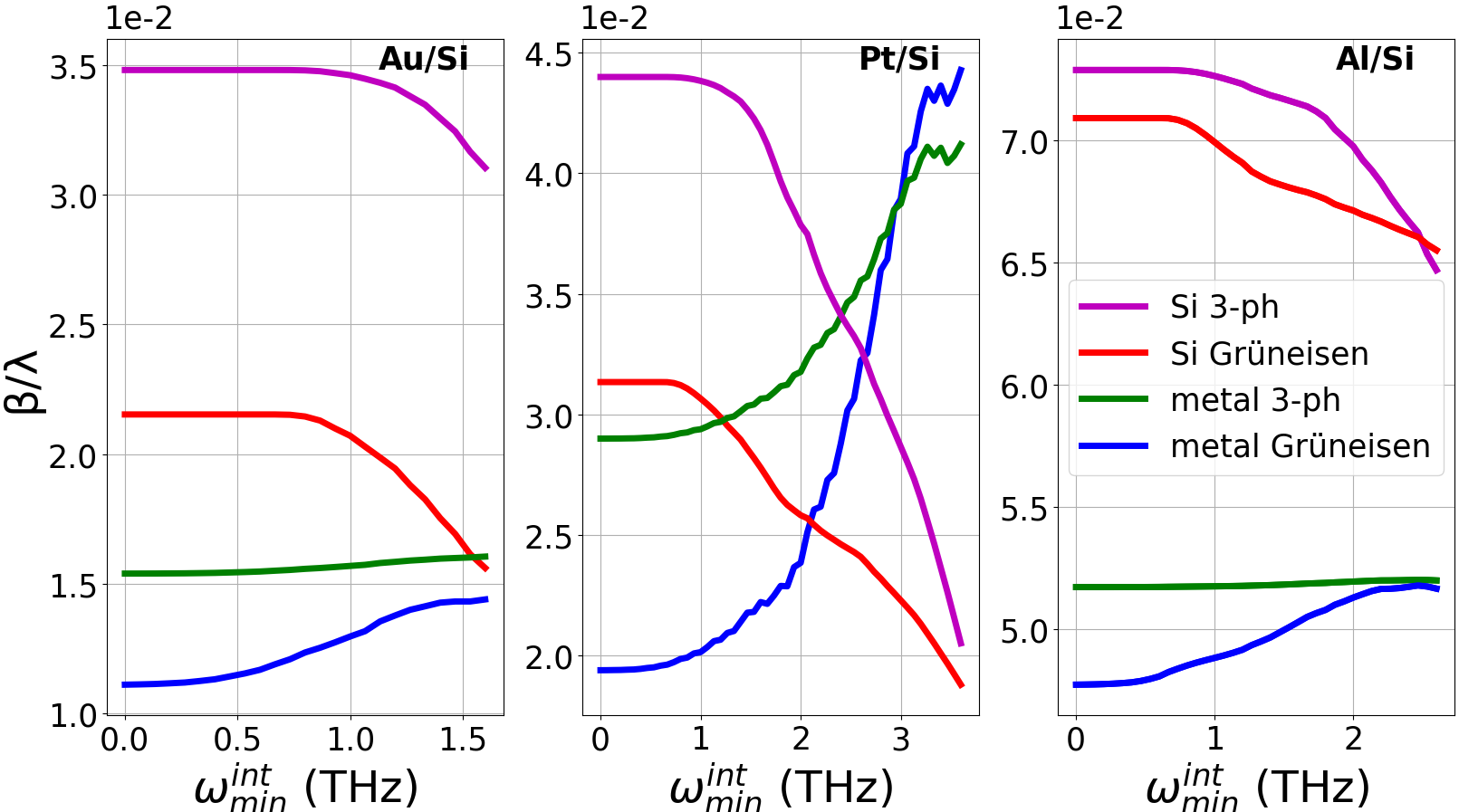}
  \caption{
  Ratio measuring the respective weight $\beta_i/\lambda_i$ of the two leads in the non-equilibrium  corrections as function of the minimal integrated frequency using either the Grüneisen parameter or the $3$ phonon models (see Eq.~\ref{eq:b_l}).} \label{fig:bg_comp}
\end{figure}

We now compare the three different approaches to compute the non-equilibrium corrections. Figure~\ref{fig:ne_corr} 
demontrates the necessity to estimate the terms $\beta_i$ through the calculations of three-phonon scattering. Indeed, the two other approaches -- although simpler -- yield non negligible differences. For the Au/Si and Pt/Si systems, the correction is greatly underestimated by both the Callaway and Grüneisen models by around $60$$\%$. One should note that the Al/Si system is a bit particular as the three approaches predict close values. There is, however, a general tendency that the Callaway and Grüneisen formalisms underestimate the deviation for all the systems. This trend can be explained by quantifying the different terms that contribute to the correction. In particular, if we consider Eq.~\eqref{eq:flux_correc}, we see that one term controls the metal contact contribution ($\beta_1/\lambda_1$) and the second the silicon contact contribution ($\beta_2/\lambda_2$). Figure~\ref{fig:bg_comp} shows the relative weight of these deviations through the ratio $\beta_i/\lambda_i$ and we can see that the main contribution to the conductance deviation stems from silicon. 
The underestimation of the correction by the Grüneisen model is explained by the fact that this model overestimates the spectral silicon mean free path, see Fig.~\ref{fig:mfp_ml}. 

\begin{figure}[h]
  \centering
  \includegraphics[width = 1.0\linewidth]{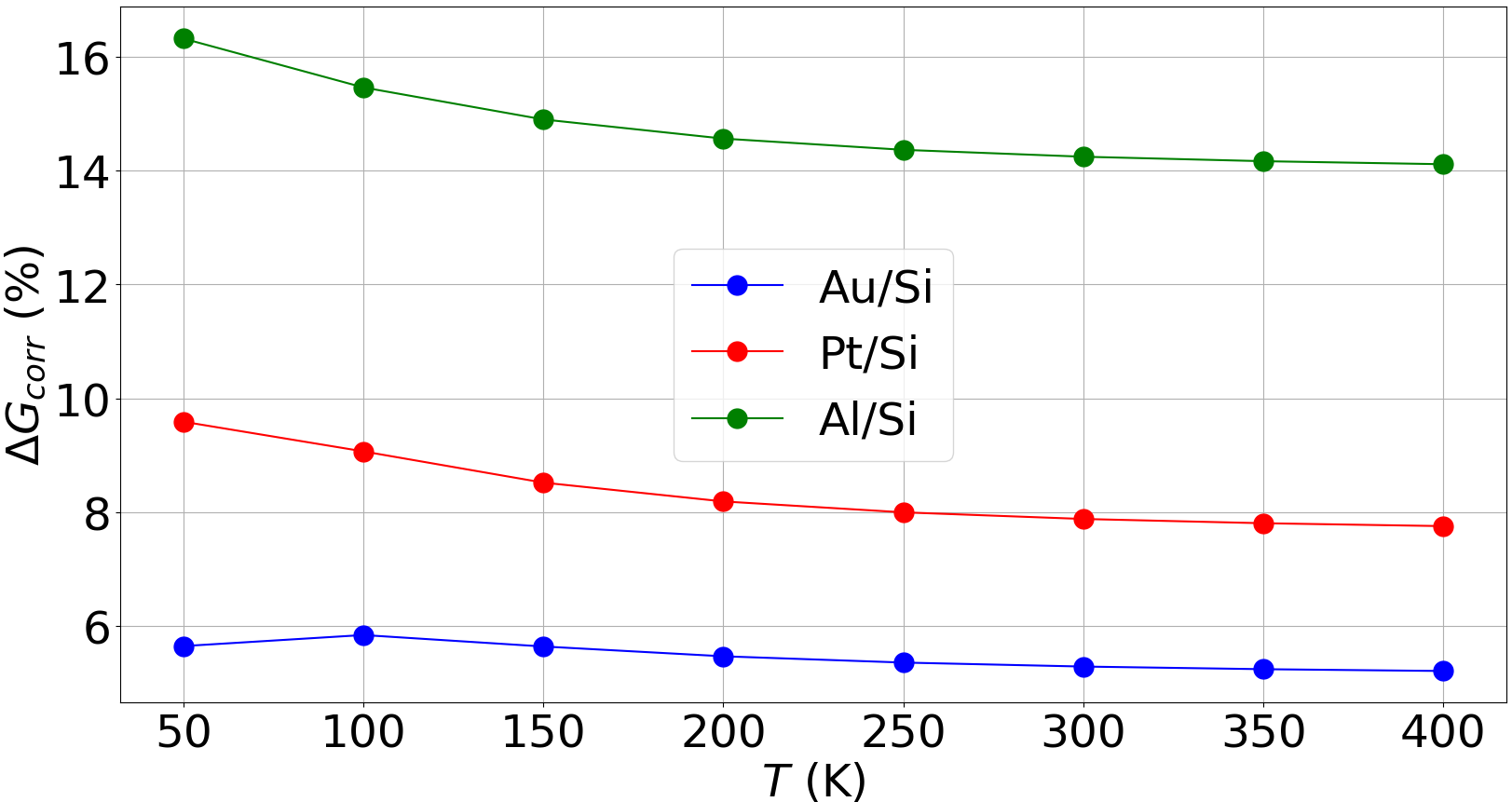}
  \caption{Relative variation of the conductance $\Delta G$ due to non-equilibrium corrections, as a function of temperature, for the three interface considered. 
  }\label{fig:ne_temp}
\end{figure}

Finally, we examine the effect of temperature on the non-equilibrium corrections. Figure~\ref{fig:ne_temp} displays the variation of the conductance with temperature for the 3 phonon approach. For all the systems considered, the corrections tend to decrease with temperature. However, the order of magnitude of the deviations remained the same over a wide range of temperature. The largest variation is observed for Pt/Si with a difference of the order of $20 \%$ between $0\,$K and $400\,$K.

\subsection{\label{sec:comparison}Comparison with experiments}
\begin{figure}
\includegraphics[width = 1.0\linewidth]{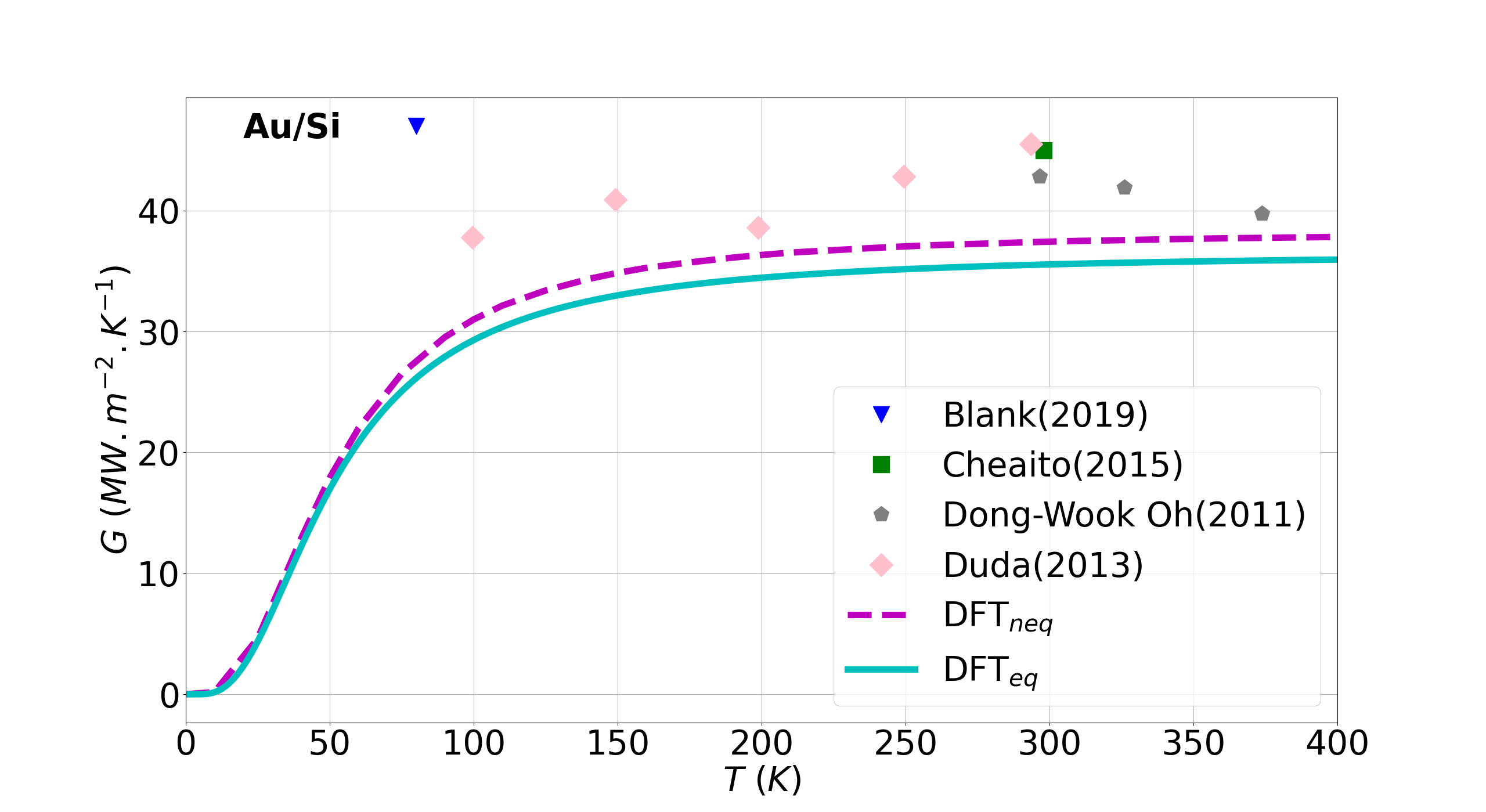}
\includegraphics[width = 1.0\linewidth]{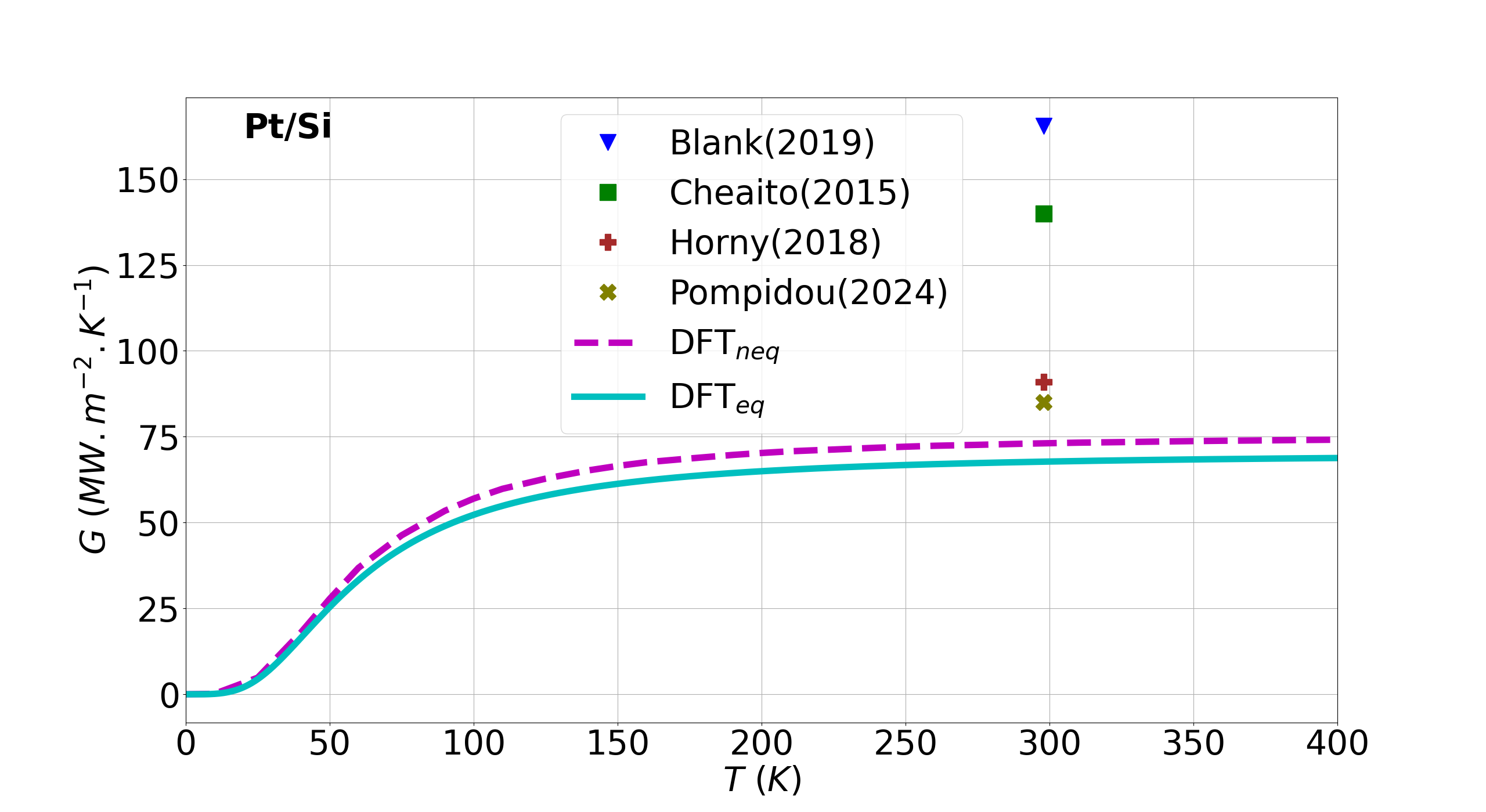}
\includegraphics[width = 1.0\linewidth]{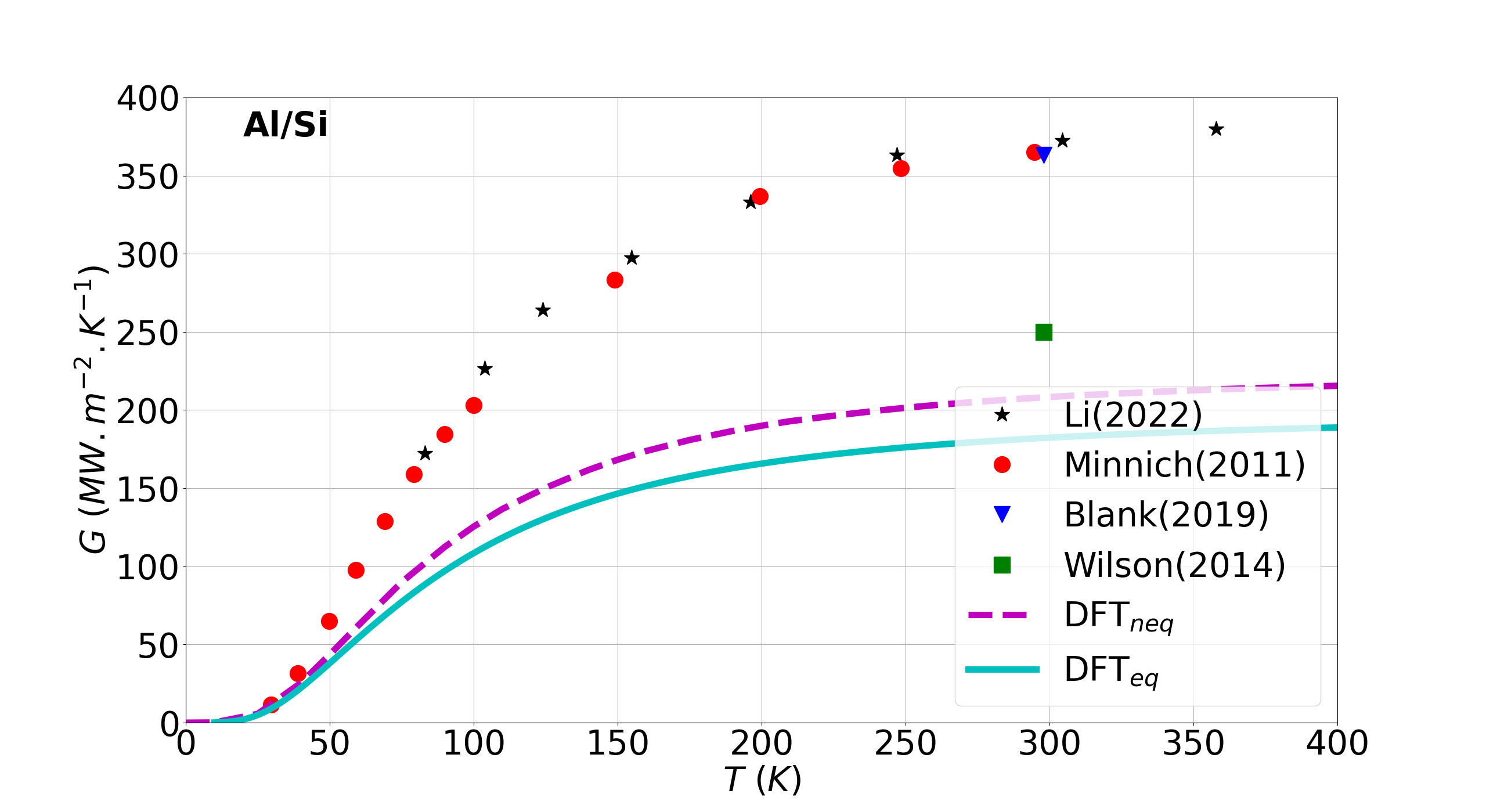}
\caption{\label{fig:ne_all} Comparison between the NEGF-DFT predictions of the ITC without (solid line) and with non-equilibrium corrections (dashed lines) compared with experimental data, for the three systems of interest, Au-Si, Pt-Si and Al-Si interfaces. The experimental data are extracted from \cite{interface_ref,Horny,
Blank2019,Dong-Wook2011,alsi_ref1,alsi_ref1,wilson2014}.}
\end{figure}

We now compare the conductance predicted by the corrected NEGF calculations to available experimental data, for the three metal-Si interfaces of interest here. Figure~\ref{fig:ne_all} shows a comparison of our ab-initio calculations with experimental data. 
In the case of the Au/Si interface, the results show that our corrected ab-initio calculations are relatively close (within $20 \%$) to the experimental results ~\cite{ausi_ref,interface_ref}. As the calculations account only for phonon-phonon processes, we conclude that energy-phonon processes turn out to have a negligible impact on interfacial heat transfer at this interface.
For the Pt-Si interface, the values predicted by ab-initio calculations are close to the experimental 
results of Horny and co-workers~\cite{Horny}, but they tend to underestimate the experimental results of Cheaito \etal~\cite{interface_ref}. For this interface, more data are clearly needed to conclude on the relative importance of electron-phonon processes.

The case of the Al/Si interface suggests a very different conclusion. In particular, we note an appreciable difference between our ab-initio calculations and the experimental data of Minnich \etal~\cite{alsi_ref2} and of Li \etal~\cite{alsi_ref1}. These experimental investigations are of special interest as they consider epitaxial systems and are therefore the most relevant to compare to our DFT calculations where the interface are atomistically smooth. 

\begin{figure}
    \centering
    \includegraphics[width=0.45\textwidth]{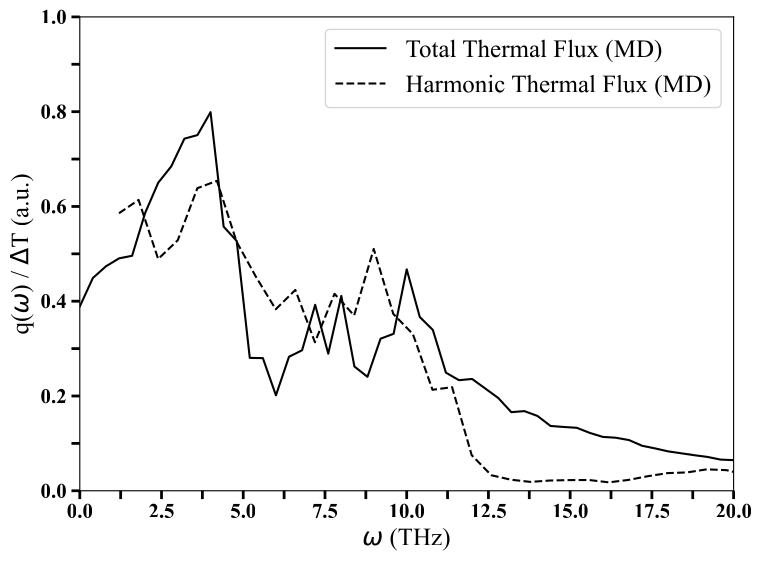}
    \caption{
    Frequency dependent total (solid line) and elastic (dashed line) thermal fluxes across the aluminum-silicon interface obtained from NEMD simulations at $300$ K, see supplementary material for details.
    }
    \label{fig:thermal_spectrum_MD}
\end{figure}

 There are two possible effects that may explain the relative differences between the first-principle calculations and the thermoreflectance data. One possible explanation stems from anharmonic phonon-phonon processes at the interface that are discarded in the NEGF calculations. 
 \corf{Anharmonic effects are generally believed to have a moderate impact on interfacial heat transfer~\cite{guo2021}. 
 
 To estimate the importance of anharmonic processes for the Al/Si interface, we have run additional non-equilibrium molecular dynamics (MD) simulations, as detailed in the supplementary material.
 Embedded Atom Model and Stillinger-Weber potentials have been employed to describe bulk Al and Si respectively, while the cross-interaction between Al and Si has been described by a Lennard-Jones potential which parameters have been tuned against NEGF calculations. 
 From these MD simulations, we extract a value of the total ITC~: (ITC$_{\mathrm{total}}^{100\,\mathrm{K}} = 154 \pm 6~\mathrm{MW\,m^{-2}\,K^{-1}}$ at $100$ K and (ITC$_{\mathrm{total}}^{300\,\mathrm{K}} = 232 \pm 24~\mathrm{MW\,m^{-2}\,K^{-1}}$ at $300$ K. These values are, respectively $9$ and $12 \%$ higher than the values of the ITC predicted by NEGF at the same temperatures. This demonstrates that anharmonic phonon-phonon effects at the interface contribute only mildly to the total thermal conductance, even at room temperature. The moderate effect of anharmonic effects is further confirmed by the analysis of the MD interface thermal spectrum $q(\omega)$ represented in Fig.~\ref{fig:thermal_spectrum_MD},
 which shows that the total thermal spectrum is well approximated by the harmonic thermal spectrum.
 In conclusion, interfacial anharmonic effects can not be invoked to explain the difference between ab-initio calculations and experimental data seen in Fig.~\ref{fig:ne_all}.}
 
 In conclusion, We interpret this difference as a signature of electron-phonon processes, the magnitude of which we will estimate in the following.
 
\subsection{Effect of the electron-phonon processes}
In this section, we discuss the effect of the electron-phonon processes on interfacial heat transfer at the metal-silicon interfaces. 

\subsubsection{Effect of electron-phonon coupling at the interfaces}
We first discuss the effect of the different electron-phonon couplings on interfacial heat transfer. At this point, it is important to distinguish electron-phonon coupling taking place in the metal and electron-phonon coupling taking place across the interface, \ie{}  a metal electron being directly coupled to a phonon in the semiconductor.
A popular approach has been proposed in 2005 by Reddy and Majumdar~\cite{majumdar2004} to model electron-phonon effects on interfacial heat transfer. The electron-phonon coupling considered in their approach is only that taking place in the bulk metal and the resulting effect turns out to be resistive. More specifically, the {\em effective} thermal conductance at metal-semiconductor interface is:
\begin{equation}
\Geff = \frac{\Gpp}{1+\frac{2 \Gpp}{\sqrt{\kp G}}},
\label{eq:Reddy_Majumdar}
\end{equation}
where $\Gpp$ is the phonon-phonon interfacial thermal conductance, $\kp$ the phonon thermal conductivity of the metal (\ie{} $\kp=k-\ke$ where $k$ is the total metal conductivity and $\ke$ the electronic thermal conductivity) and $G$ the electron-phonon coupling constant. From this latter expression, it is straightforward to see that $\Geff<\Gpp$, \ie{} electron-phonon processes tend to reduce the effective thermal conductance $\Geff$. 

It is, however, important to remark that the above treatment of the effect of electron-phonon processes is valid only in the absence of a direct electron-phonon coupling, that is $\Gep$ should be strictly vanishing. In the following, we will demonstrate from ab-initio calculations that there exists a direct coupling between metal electrons and silicon phonons. Therefore, the adiabatic condition for the electrons at the interface assumed in the Reddy and Majumdar model does not apply generally.

In Appendix~A, we derive the expression of $\Geff$ in the presence of a non vanishing conductance $\Gep$. It is shown that, to a good approximation, $\Geff$ is given by:
\begin{equation}
\Geff = \Gpp + \Gep-\Gpp^2\frac{\sqrt{\kp}}{G}. 
\label{eq:Geff_steady_state}
\end{equation}
This expression shows that $\Geff$ may be larger than $\Gpp$. 
However, we will not employ this expression to infer $\Gep$ from the experimental values $G_{\rm exp}$ as thermoreflectance experiments do not correspond to steady state heat flow situation, but rather to transient thermal relaxation. Before going further and investigating the solution of the two-temperature model (TTM) under transient conditions, we will show evidence from ab-initio calculations that a non vanishing interfacial electron-phonon coupling takes place at the interface. 

\subsubsection{Evidence of a direct electron-phonon coupling}
Here, we discuss the evidence we have for the existence of a direct electron-phonon coupling. To start with, we consider the expression of the interfacial electron-phonon conductance:
\begin{equation}\label{eq:g_ep}
    \Gep = \frac{2 k_B N_f L}{V}\int_0^{\infty}\alpha^2 F(\omega) \hbar \omega 
 d\omega,
\end{equation}	
where $\alpha^2 F(\omega)$ is the Eliashberg function, $N_f/V$ the electronic density of state per unit of volume and $L$ the width of the system along the thermal transport direction. The Eliashberg function is defined as:
\begin{equation}\label{eq:eliash}
    \alpha^2 F(\omega) = \frac{1}{2\pi N_f} \sum_{{\bm q}\nu} \frac{\Pi_{{\bm q}\nu}}{\omega_{{\bm q}\nu}} \delta(\hbar\omega - \hbar \omega_{{\bm q}\nu}).
\end{equation}	
$\Pi_{{\bm q}\nu}$ corresponds to the phonon linewidth:
\begin{equation}
\begin{aligned}
    \Pi_{{\bm q} \nu} = &2 \pi \omega_{{\bm q}\nu} \sum_{{\bm k} \rm {mn}} |g_{\rm {mn},\nu}({\bm k},{\bm q})|^2 \\
    &\times \delta (\hbar \omega_{{\bm q}\nu} + E_{\rm n {\bm k}} - E_{\rm m {\bm k+q}}), 
    \label{eq:pi_qnu}
\end{aligned}
\end{equation}	
where $E_{n{\bm k}}$ and $E_{m{\bm k}+{\bm q}}$ are eigenvalues (eigen energies) of electronic vectors $n{\bm k}$ and $m{\bm k}+{\bm q}$ with $n$ and $m$ the considered electronic modes. $E_f$ is the Fermi energy and $g_{\rm {mn},\nu}({\bm k},{\bm q})$ are the electron-phonon matrix elements. 

The electron-phonon matrix elements $g_{\rm {mn},\nu}({\bm k},{\bm q})$ are expressed as:
\begin{equation}\label{eq:gmnk}
\begin{aligned}
    &g_{\rm {mn},\nu}({\bm k},{\bm q}) = \langle \Psi_{\textit{n}{\bm k}}|\partial_{{\bm q} \nu}V|\Psi_{\textit{m}{\bm k}+{\bm q}} \rangle,
\end{aligned}
\end{equation}
where $\Psi_{\textit{n}{\bm k}}$ and $\Psi_{\textit{m}{\bm k}}$ are the electronic wavevectors. 




\begin{figure}[h]
\includegraphics[width = 1.0\linewidth]{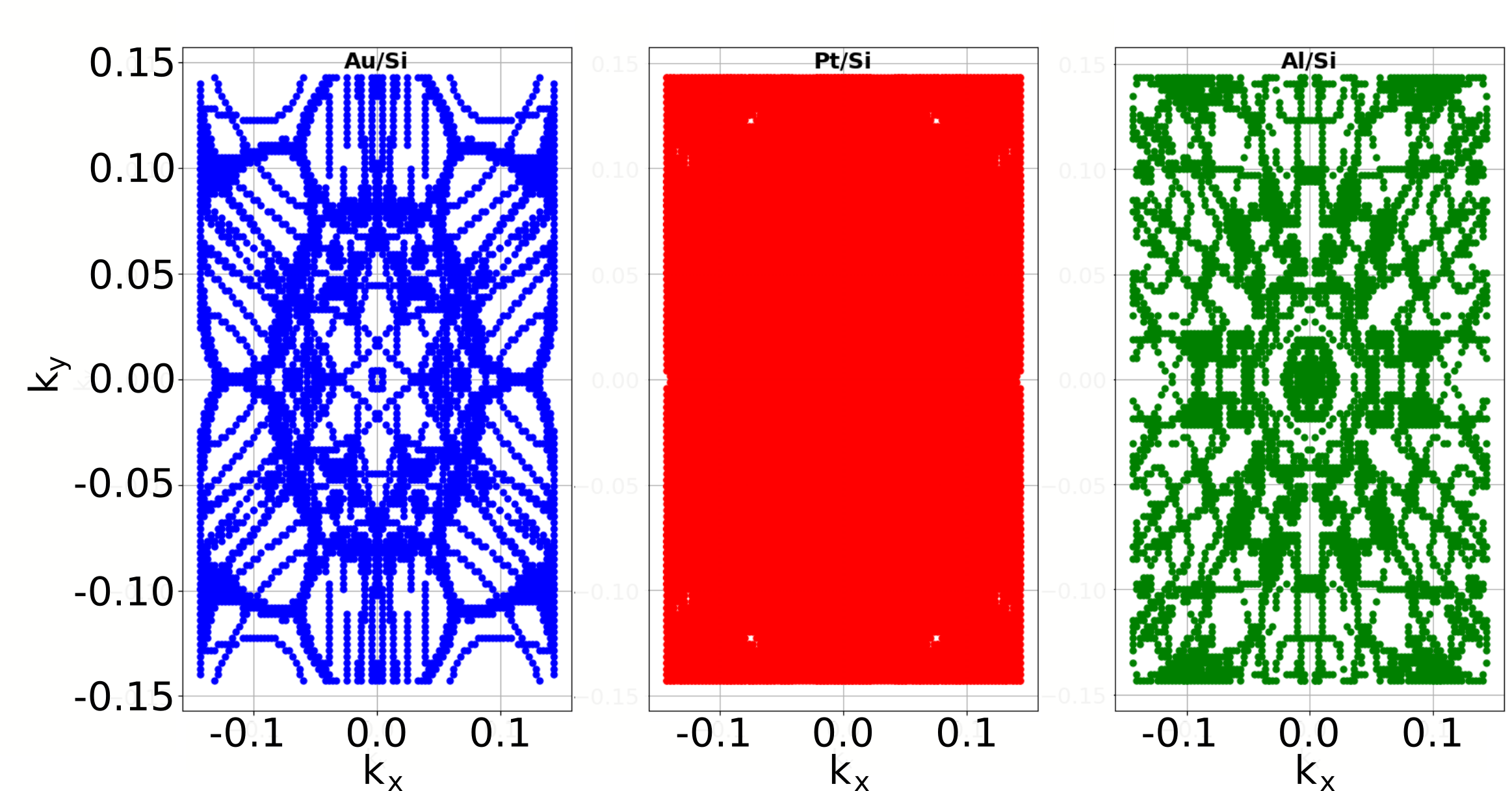}
\caption{Fermi surface calculated for Au/Si, Pt/Si and Al/Si interfaces with a reciprocal space mesh of $k_x\times k_y\times k_z = 100 \times 100 \times 1$. $k_z$ corresponds to the direction of thermal transport. The tolerance threshold around the Fermi energy is here $\sigma_{E_F} = 10$ meV.\label{fig:fermi_surf_10mev}}
\end{figure}

The full calculation of the electron-phonon conductance 
(Eqs.~\eqref{eq:g_ep}-\eqref{eq:pi_qnu})  
is currently too heavy to be implemented numerically. For this reason, we have rather estimated the different terms that appear in Eq.~\eqref{eq:g_ep} for the three scrutinized systems. 

We first show evidence that the interfacial electron-phonon conductance $\Gep$ is non vanishing. To this end, it is sufficient to show that the discrete sum over the electronic modes in Eq.~\eqref{eq:pi_qnu} is not restricted to zero, a fact that is confirmed by the analysis of the phonon density of states of the different interfaces represented in Fig.~\ref{fig:all_pdos}.
To proceed, we approximated the term $\delta (\hbar \omega_{\bm q\nu} + E_{\rm nk} - E_{\rm mk+q})$ of Eq.~\eqref{eq:pi_qnu} which relates the energy difference between two electronic modes and a phonon. The approximation follows two selection rules:
    1) only electron modes $k$ with $|E_{k}-E_F|\leq\sigma_{E_F}$ are selected (with $\sigma_{E_F}$ the maximum deviation energy accepted around the Fermi level). The electronic vectors of these electrons allow one to construct the set of vectors: $\Delta \textit{nm}{\bm k} = m{\bm k} - n{\bm k}.$ \\
%
%
    2) Only the phonon modes $q$ of vector $q_{\Delta \rm{nm}{\bm k}} = \Delta \rm{nm}{\bm k}$ and such that $|E_{{\bm q}_{\Delta \rm{nm}{\bm k}}}-E_{\Delta \textit{nm}{\bm k}}| \leq \sigma_{E_F}$ are selected with: $E_{{\bm q}_{\Delta  \rm{nm}{\bm k}}} = \hbar \omega_{{\bm q}_{\Delta  \rm{nm}{\bm k}}}$.
%
The selection of electronic modes is performed by constructing Fermi surfaces. 
From now on, we will rather use the wording "Fermi contours" because we explore the Brillouin zone only in the directions orthogonal to the thermal transport axis.
These contours were calculated on a Brillouin zone divided into $100 \times 100$ elements and a tolerance $\sigma_{E_F} = 10$ meV much smaller than the Fermi energy $E_f$ of the systems under consideration. 

The Fermi contours, presented in Fig.~\ref{fig:fermi_surf_10mev}, 
yield an important conclusion: 
The set of electronic transition energies $\Delta \textit{nm}{\bm k}$ is sufficiently dense so that the set of phonons $q_{\Delta \textit{nm}{\bm k}}$ constructed enables a complete sampling of the Brillouin zone for phonon modes having an energy lower than  $2\sigma_{E_F}$. This means that the calculation of the phonon eigenmodes of the set of vectors ${\bm q}_{\Delta \textit{nm}{\bm k}}$ will lead to a density of states very similar to the phononic density of states of Fig.~\ref{fig:all_pdos} for frequencies lower than $\omega_c=\sigma_{E_F}/\hbar \simeq5$ THz. 

We conclude from this analysis that the number of electron-phonon channels in Eq.~\eqref{eq:pi_qnu} is large and there is therefore a non vanishing coupling $\Gep$ for the three systems under investigation.

\subsubsection{Electron-phonon interfacial thermal conductance}
Now that we have demonstrated that a non vanishing electron-phonon coupling takes place at the the interface, we will estimate the corresponding electron-phonon interface thermal conductance $\Gep$.
To this end, we will employ the numerical solution of the two-temperature model~(TTM). All the quantities that appear in the TTM are calculated from ab-initio calculations or taken from the literature. The only unknown is the electron-phonon conductance $\Gep$ that is treated as a free parameter, and which value will be inferred from the comparison between the numerical solution of the TTM model and the effective thermal conductance $\Geff$ measured experimentally by time-domain thermoreflectance.

To model a situation which is relevant to the thermoreflectance experiments, we consider a thin metal film of thickness $h$ lying on a silicon substrate. A laser pulse heats up the metal electrons and the energy supplied by the laser is absorbed over a penetration depth $\delta$ smaller than the film thickness. The values of the phonon-phonon conductance $\Gpp$ are taken from the ab-initio calculations presented in Fig.~\ref{fig:ne_all}, namely $\Gpp=38, 65$ and $210$ MW/m$^2$/K for gold,platinum and aluminium-silicon interfaces respectively. The other metal and silicon properties are taken from Tabs.~\ref{Table_parameters} and \ref{Table_thermo_parameters}.  

\begin{table}[h]

\begin{tabular}{lrrrr}
\hline 
\hline 
 metal  &  Au  &  Pt & Al  & units            \\
 \hline
 $G$  &  26   & 1090  &  246  &  \hspace{1mm} 10$^{15}$ W/m$^3$/K           \\
 $\kp$ &  2.8  &   6.49 &  8.95  & W/m/K  \\
 $k$          &   318  &    71.9 &   237 & W/m/K    \\
$\lep$  & 10  & 2.44  &   1.65 & nm             \\
$\kp/\lep$ &   280  &  2650 &   5420 & MW/m$^2$/K  \\
\hline \hline
\end{tabular}

\caption{Electron-phonon parameters at $300$ K: 
electron-phonon coupling constant $G$,
phonon thermal conductivity~$\kp$,
thermal conductivity $k=\kp+\ke$,
electron-phonon length~$\lep$.
The values of $G$ are taken from Ref.~\cite{lin2008}, 
while the values of $k$ and $\ke$ are taken from Ref.~\cite{tong2019}.  \label{Table_parameters}}
\end{table}

\begin{table}[h]
\begin{tabular}{lrrrrr}
\hline \hline  
 medium  & Au & Pt & Al & Si & units \\ 
 \hline
 $\gamma$   &  71.4  & 741  & 136.3 & -& J/m$^3$/K$^2$ \\
 $C_{\rm p}$ & 2.42  & 2.85 &4.1  & 1.63 & \hspace{1mm}$10^6$ J/m$^3$/K \\
 $\rho_{\rm m}$  & 19300 & 21400 &  2720  & 2330 & kg/m$^3$ \\
\hline  \hline
\end{tabular}
\caption{Thermophysical parameters at $300$ K: 
Sommerfeld constant $\gamma$,
specific heat $C_{\rm p}$ and
mass density $\rho_{\rm m}$. These values are extracted from Ref.~\cite{kittel}}.
\label{Table_thermo_parameters}
\end{table}



In the following, we have taken typical values $h=100$ nm and $\delta=20$ nm.
The effective thermal conductance is determined based on the time it takes for phonons to see their temperature decrease by half their maximum temperature. 
Further details regarding the procedure and the numerical approach may be found in Appendix B.
Specifically, we take  for the experimental values of the thermal conductance at room temperature~:~$\bcexp =43$  and 350  \bcunit{} for Au/Si and Al/Si interfaces respectively, and $\bcexp$  in the interval [80-150]  \bcunit{} for the Pt/Si interface.  
The values of the interfacial electron-phonon conductance deduced through this procedure are found to be:
\begin{subequations}
\begin{align}
\mbox{for Au/Si}\qquad \bce&= 5    \qquad\:\:\:\:  \mbox{\bcunit},  \label{eq:GepfromTTMa} \\
\mbox{for Pt/Si}\qquad \bce&= 15-80\,      \mbox{\bcunit},  \label{eq:GepfromTTMb} \\
\mbox{for Al/Si}\qquad \bce&= 107  \quad\:\:\:\:  \mbox{\bcunit}.  \label{eq:GepfromTTMc}
\end{align}
\end{subequations}
These values are in lines with our previous analysis of the importance of electron-phonon processes. For Au/Si, the interfacial electron-phonon conductance amounts to $12 \%$ only of the total conductance, confirming that for this system, electron-phonon processes play a minor role. For Pt/Si, the dispersion of the experimental data in the literature yields an interval of values for $\bce$. Further experimental data are clearly needed to draw conclusions on the importance of electron-phonon effects in this system. 
Finally, for the Al/Si system, we clearly see that the electron-phonon interfacial conductance represents around one third of the total conductance. In this system, interfacial electron-phonon processes can not be neglected.

\subsubsection{Amplitude of the derivative of the deformation potential}

To better understand the physical origin of the interfacial electron-phonon conductance, we analyze the derivative of the perturbation potential $\partial V_{{\bm q}\nu}(r)$. 
This term is one of the main contribution of the electron-phonon matrix elements $g_{\rm {mn},\nu}$ (see Eq.~\eqref{eq:gmnk}) and follows directly from the electron-ion and electron-electron interaction potentials. 

\begin{figure}[h]
\includegraphics[width = 1.0\linewidth]{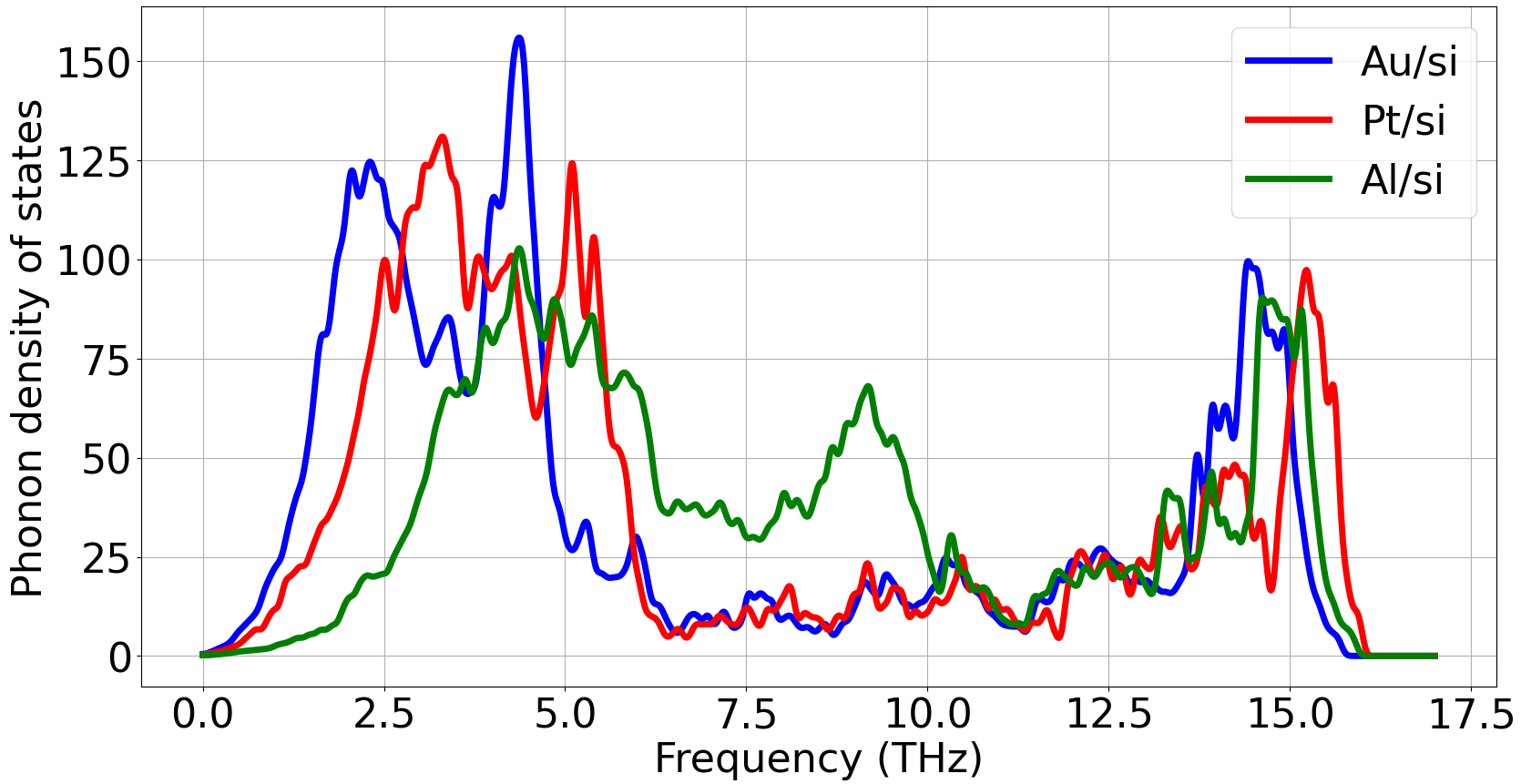}
\caption{Phonon density of states for Au/Si(blue), Pt/Si(red) and Al/Si(green) interfaces obtained using phonopy and optimized ab-initio systems.\label{fig:all_pdos}}
\end{figure}

To interpret the spatial variations of the deformation potential $\partial V_{{\bm q}\nu}(r)$, it is first important to analyze the phonon density of states of the three systems shown in Fig.~\ref{fig:all_pdos}. First, focusing on the two systems Au/Si and Pt/Si, we can distinguish two regions: one at low frequencies corresponding to the acoustic modes of the metals below $6$ THz and the second at high frequencies corresponding to the optical modes of silicon above $13$ THz. For Al/Si, we have an excess of intermediate modes that correspond to the acoustic modes of Al below $11$ THZ and again another excess corresponding to the optical modes of silicon above $13$ THz. 

Now, let us focus on the analysis of the deformation potential $\partial V_{q\nu}(r)$
:~this term can be calculated from the phonon perturbation potential $V_{R_a,R_p}$ build by SIESTA for the supercell $3\times3\times1$, where $R_a$ is the atomic position of atom $a$ in the primitive cell and $R_p$ the lattice vector to translate the primitive cell to the $p$-th cell of the supercell system. For one atom and one perturbative displacement $\eta_{\alpha}$ ($\alpha$ corresponding to one of the three bulk crystal axis), we express the derivative as:
\begin{equation}\label{eq:d_vn_vp}
    \partial V_{R_a,R_p,\alpha} = \frac{V_{R_a+\eta_{\alpha},R_p}-V_{R_a,R_p}}{\eta_{\alpha}}.
\end{equation}	

\begin{figure}[h]
\includegraphics[width = 1.0\linewidth]{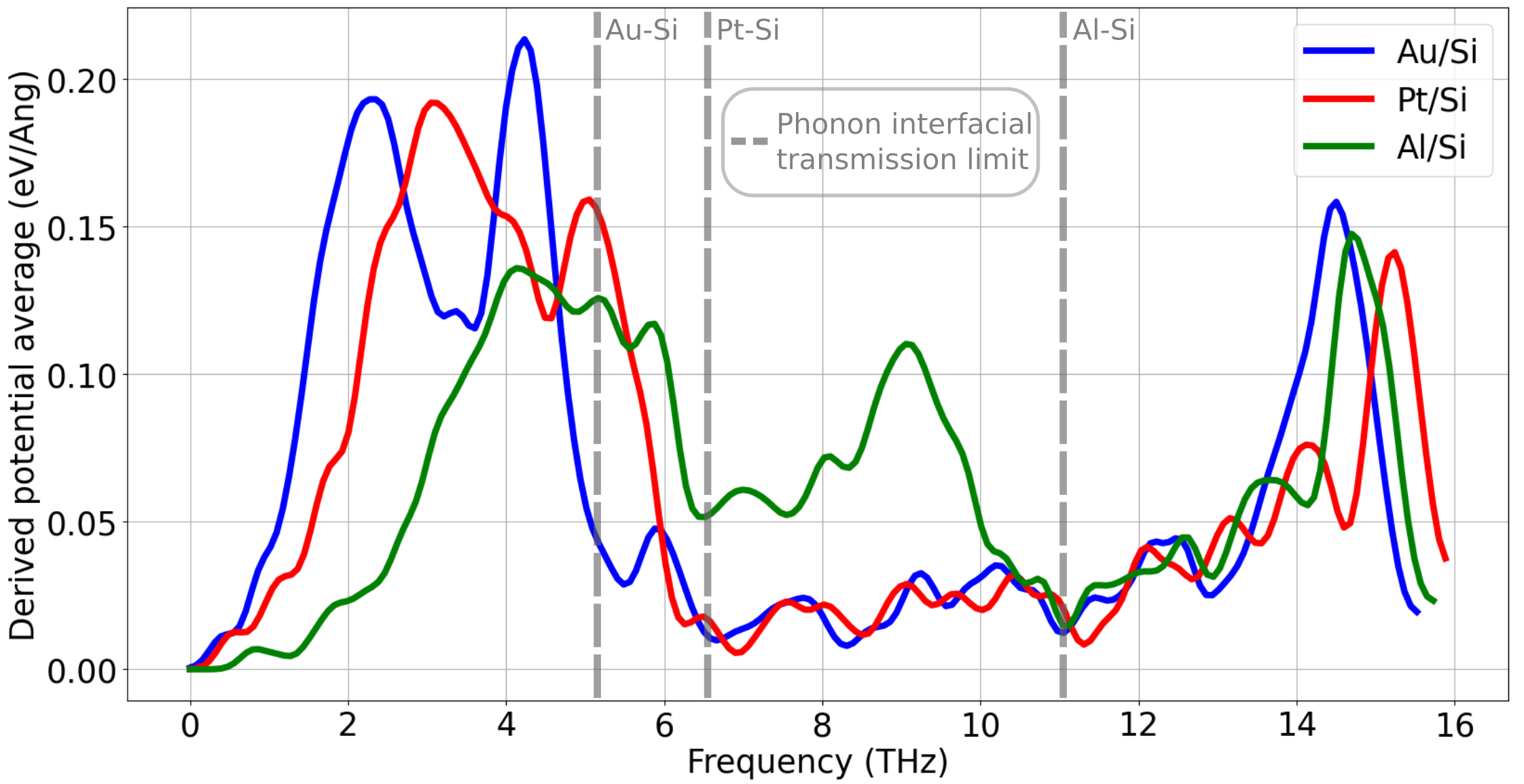}
\caption{Average amplitude of the derivative of the deformation potential $\partial V_{q\nu}$ for the three interfaces Al/Si (green), Pt/Si (blue) and Au/Si (red) calculated at the special point $K$. The dotted lines correspond to the frequency limit of the phonon transmissions shown in Fig.~\ref{fig:trans}. \label{fig:meanpot}}
\end{figure}

The derivative of the phonon perturbation can be defined as:
\begin{equation}\label{eq:d_vqnu}
    \partial V_{\bm q\nu} = \sum_{{\bm R}_a,{\bm R}_p,\alpha} \partial V_{{\bm R}_a,{\bm R}_p,\alpha} e_{\bm q\nu}^{{\bm R}_a,\alpha} e^{i {\bm q}\cdot {{\bm R}_p}}.
\end{equation}	
Since the computation of $\partial V_{\bm q\nu}(r)$ for a large range of phonon modes $q\nu$ is heavy and expensive, we chose to focus on the special point $K$ as it corresponds to the coordinate $k_z =0$, that is perpendicular to the direction of thermal transport. 
We focus on the spatial average (along the three directions) of the derivative of the deformation potential shown in Fig.~\ref{fig:meanpot}. The deformation potential tends to follow the spectral dependence of the phonon density of states, as shown in Fig.~\ref{fig:all_pdos} 
and we can not evidence a large difference between the three systems. In particular, the Al/Si system does not seem to display a larger deformation potential as compared to Au/Si.

From Fig.~\ref{fig:meanpot}, we deduce that any difference between the three systems should 
stem from the different range of phonons participating in interfacial transfer, as already calculated in Fig.~\ref{fig:trans}. Based on this latter figure, it is reasonable to assume 
that only modes having a frequency smaller than the cut-off frequency of the metal participate in the interfacial electron-phonon coupling.
This implies that only phonon modes involved in phonon-phonon processes participate to electron-phonon mediated interfacial heat transfer. Therefore, electron-phonon coupling at the interface proceeds along two paths:


    1) an electron-phonon interaction taking place within the metal alone
 
    2) phonon-phonon interaction between the phonons of silicon and phonons from the electron-phonon interaction of the metal.

\begin{figure}[h]
\includegraphics[width = 1.0\linewidth]{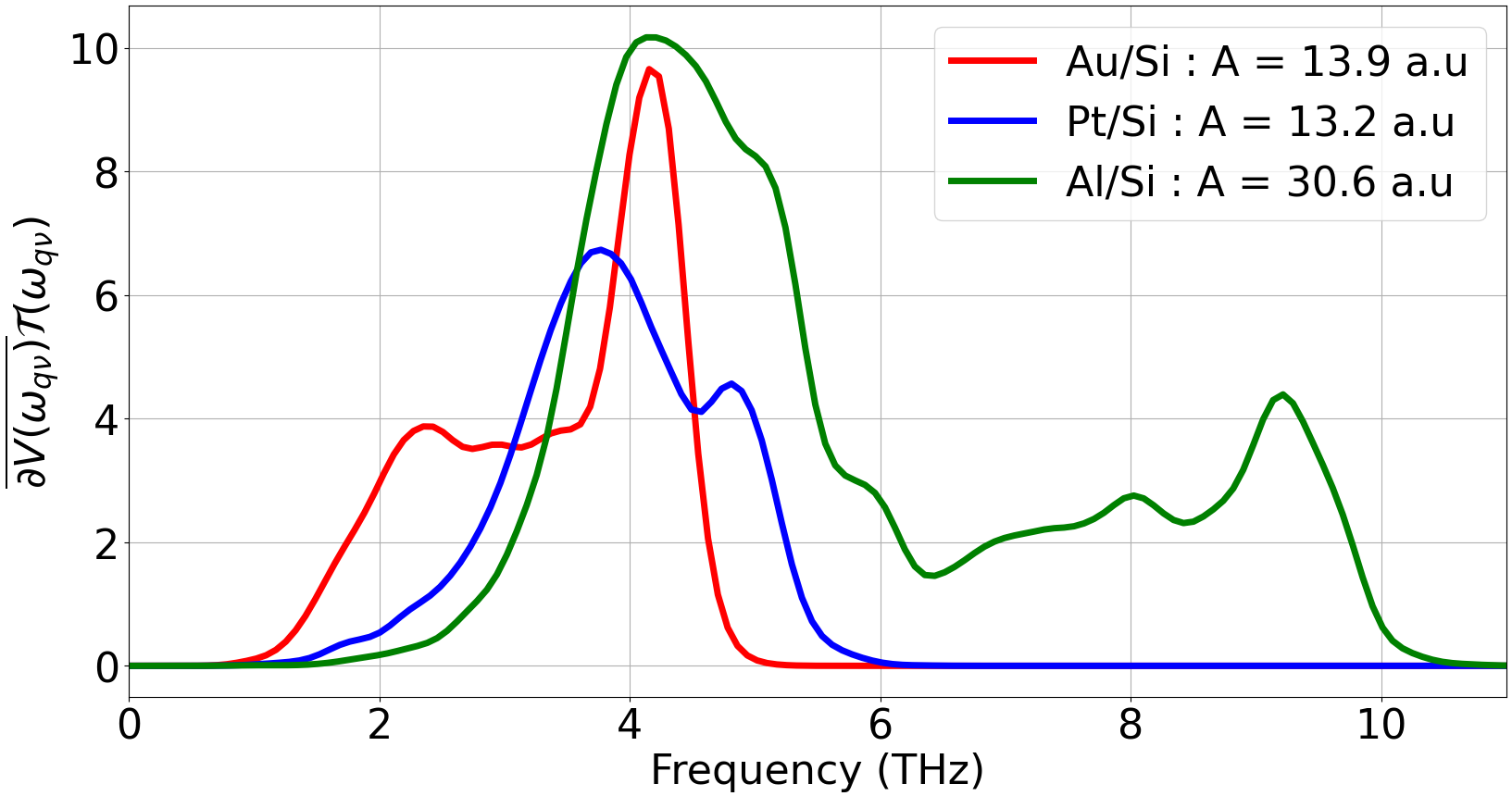}
\caption{Average amplitude of the derivative of the deformation potential $\partial V_{q\nu}$ multiplied by the phononic transmission $\mathcal{T}(\omega)$ for the three interfaces Al/Si (green), Pt/Si (blue) and Au/Si (red) calculated at the special point $K$. The quantity $A$ indicated in the legend corresponds to the area under the curve for the different interfaces. \label{fig:mpot}}
\end{figure}

To be specific, we have represented in Fig.~\ref{fig:mpot} the spectral deformation potential multiplied by the phonon transmission shown in Fig.~\ref{fig:trans}.

In Fig.~\ref{fig:mpot}, the cumulative deformation potential characterizing the Al/Si interface is the largest among the three systems. This difference in amplitude is essentially explained by a broader phonon spectrum in the case of Al/Si as compared to Au/Si and Pt/Si. Figure~\ref{fig:mpot} shows that in the case of indirect electron-phonon coupling at the interface, there is a amplitude difference depending on the study system, which could explain the differences observed between simulations and experiments for the Au/Si and Al/Si interfaces.

\section{\label{sec:conclusion}Conclusions}
The objective of this study was to quantify the role of electron-phonon processes on interfacial heat transfer at metal-silicon interfaces. To this end, we performed a combination of ab-initio calculations and non-equilibrium Green's function (NEGF). We selected three systems, namely Au-Si, Pt-Si and Al-Si, so as to span a range of acoustic contrast between metal and silicon.

We first propose a method to extend NEGF calculations to account for out-of-equilibrium effects discarded in the original formulation. We show that out-of-equilibrium effects may increase the interfacial thermal conductance by $15 \%$ with respect to the NEGF predictions. 
After applying these corrections, we compare the interfacial thermal conductance to experimental data of the literature. For Au-Si interfaces, the ab-initio calculations are in good agreement with experiments. Electron-phonon processes turn out to play a minor role at this interface. For Pt-Si interfaces, the dispersion of experimental data does not allow us to be conclusive. Further experimental data are therefore needed, especially at low temperatures to estimate the importance of electron-phonon processes. 
By contrast, we report a difference between the NEGF calculations which account only for phonon-phonon processes and experimental investigations on epitaxial Al-Si interfaces. 
We interpret these discrepancies as a signature of interfacial electron-phonon processes. Numerical solution of the two-temperature model allows us to estimate than one third of the total heat transfer across this interface is mediated by the direct electron-phonon channel.
\corf{Overall, we saw the primary role played by the metal Debye frequency. The metal Debye frequency conditions the range of frequencies of phonons transmitted at the interface. This implies that the phonon-phonon thermal conductance calculated here at the harmonic level increases with the metal’s Debye frequency. As explained in Sec.~E., this has the further consequence that interfacial electron-phonon processes tend to be stronger for metals having large Debye frequency.}

Several perspectives emerge from this study:~first, this work opens the door to the systematic comparison between the thermal conductance predicted by first-principle calculations and thermoreflectance data so as to infer the value of the interfacial electron-phonon conductance based on the two-temperature model framework. Second, altough we demonstrated that anharmonic effects at the interface have a moderate effect, it would be interesting to account for them either by extending the NEGF formalism~\cite{guo2021}
or through machine-learning molecular dynamics simulations~\cite{rajabpour2025}. A third route for future works is the development of approximate approaches to estimate the electron-phonon interfacial conductance based on ab-initio calculations so as to identify which vibrational modes are coupled with the metal electrons. Finally, all the knowledge gained in the computation of interfacial electron-phonon processes may exploited to optimize the thermal response of heterogeneous metallic nanoparticles for thermoplasmonic applications~\cite{alkurdi2020,elhajj2025}.  








\section*{Appendix A: Derivation of the effective thermal conductance under steady state conditions}
\label{sec:appendix_A}

\begin{figure}[b!]
\includegraphics[width = 0.7\linewidth]{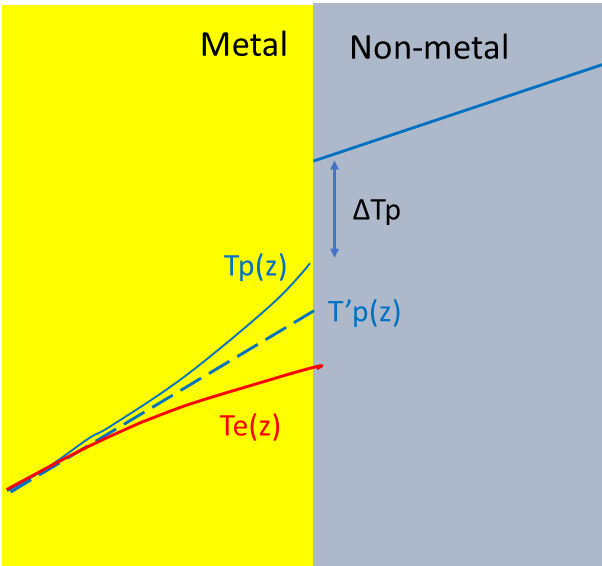}
\caption{Situation considered for the calculation of the effective thermal conductance under steady state conditions.\label{fig:interface_elec_phonon}}
\end{figure}

The situation we consider is illustrated in Fig.~\ref{fig:interface_elec_phonon}. 
The steady state equations obeyed by the electronic and phonon temperatures in the metal, denoted respectively $\Te$ and $\Tp$, write:
\begin{subequations}    
\begin{align}
  \ke \frac{d^2 \Te}{dz^2} -G(\Te-\Tp) &= 0, \label{eq:steady_state_Te}\\
  \kp \frac{d^2 \Tp}{dz^2} -G(\Tp-\Te) &= 0. 
\label{eq:steady_state_Tp}
\end{align}
\end{subequations}
In particular, we obtain from this set of equations in steady state:
$\ke \frac{d^2 \Te}{dz^2}+\kp \frac{d^2 \Tp}{dz^2}=0$ implying that the total heat flux $J=-\ke \frac{d \Te}{dz}-\kp \frac{d \Tp}{dz}$ is uniform throughout the system. 
Equations~\eqref{eq:steady_state_Te} and \eqref{eq:steady_state_Tp} 
are supplemented by the following boundary conditions:
\begin{subequations}
\begin{align}
  -\ke \frac{d \Te}{dz} &= \Gep(\Te-\Ts) \quad \mathrm {for} \quad z = 0, 
  \label{eq:boundary_condition_Te} \\ 
  -\kp \frac{d \Tp}{dz} &= \Gpp(\Tp-\Ts) \quad \mathrm {for} \quad z = 0,
  \label{eq:boundary_condition_Tp}
  \end{align}
  \end{subequations}
where $\Ts(z)$ is the temperature field in the semiconductor.  
Equation~\eqref{eq:boundary_condition_Te} is different from the adiabatic boundary condition at the metal/semiconductor interface assumed in Ref.~\cite{majumdar2004},  
as required by the fact that the electron-phonon $\Gep$ interfacial conductance does not vanish.
Defining $\theta(z)=\Tp(z)-\Te(z)$, 
the solution to Eqs.~\eqref{eq:steady_state_Te}-\eqref{eq:steady_state_Tp} writes:
\begin{eqnarray}
 \theta(z) & = & \theta_0  \exp \left(z/\lep \right) \quad \mathrm {for} \quad z \le 0, 
\label{eq:solution_theta}
\end{eqnarray}
where
\begin{equation}
  \lep=\sqrt{\left( \frac{1}{\ke} + \frac{1}{\kp} \right)^{-1} G^{-1}},  
  \simeq \sqrt{\frac{\kp}{G}}
  \label{eq:electron_phonon_length}
\end{equation}
is the electron-phonon non-equilibrium length~\cite{majumdar2004}, and the last approximation holds for metals having $\kp \ll \ke$.

To solve the set of equations obeyed by both $\Tp(z)$ and $\Te(z)$, it is useful to introduce the intermediate field $S(z)=\kpbar \Tp(z)+ \kebar \Te(z)$,
with $\kpbar=\kp/k$ and $\kebar=\ke/k$, and $k$ the total thermal conductivity. The phonon and electron temperatures may be obtained from $\theta(z)$ and $S(z)$:
\begin{subequations}
\begin{align}
 \Tp(z) & =   \kebar \theta + S, \\
 \Te(z) & =  -\kpbar \theta + S,
\label{eq:solution_theta}
\end{align}
\end{subequations}
In steady state, the total heat flux $J$ is uniform. This condition allows us to express $S(z)$ from $-k\frac{dS}{dz}=J$, which can be integrated:
\begin{equation}
S(z)=-\frac{J}{k}z+S_0.
\end{equation}
The two constants $\theta_0$ and $S_0$ may be obtained from the two boundary conditions, Eqs.~\eqref{eq:boundary_condition_Te}-\eqref{eq:boundary_condition_Tp}):
\begin{eqnarray}
J &=&-(\Gpp+\Gep) \Delta_0 - (\Gep\bar{k}_{\rm p}-\Gpp\bar{k}_{\rm e}) \theta_0, \\
\frac{\theta_0}{\lep} &=& \left(\frac{\Gpp}{\kp}+\frac{\Gep}{\ke}\right) \Delta_0 - \frac{(\Gep{\kp}^2+\Gpp{\ke}^2)}{k \ke \kp} \theta_0,
\nonumber \\  
\end{eqnarray}
where $\Delta_0=T_s(0)-T_p(0)$. To compute the effective thermal boundary conductance $\Geff$, we may  
follow~\cite{majumdar2004} and define it through:
\begin{equation}
\Geff=\frac{J}{\Ts(0)-{T'}_{\rm p}(0)}, 
\label{effective_thermal_conductance}
\end{equation}
where $\Ts(z)$ is the temperature field in the semiconductor and ${T'}_{\rm p}(z)$
is the apparent phonon temperature field in the metal, see Fig.~\ref{fig:interface_elec_phonon}: 
$T'_p(z)=-(J/k)z + S_0$.
To compute $\Geff$, we need to determine $\Ts(z)$. 
In steady state, 
$\Ts(z)=Az+B$, where the two constants $A$ and $B$ are determined based on the
boundary conditions:
\begin{subequations}
\begin{align}
    -\ks \frac{d \Ts}{dz} &= J,  \\ 
    -\ks \frac{d \Ts}{dz} &= -\Gpp(\Ts-\Tp)-\Gep(\Ts-\Te),  
\label{eq:boundary_condition_semicon}
    \end{align}
    \end{subequations}
for $z=0$. 
From the first boundary condition Eq.~\eqref{eq:boundary_condition_semicon}, 
one has ~$A=-J/\ks$. 
The second constant may be determined using the second equation 
in Eq.~\eqref{eq:boundary_condition_semicon}: 
$B=J/\Gpp+\kebar \theta_0 + S_0$. 
It is now straightforward to derive the expression of the effective thermal conductance~
\begin{equation}
\Geff=\Gpp + \Gep + \frac{ \Gep\kpbar^2 - \Gpp\kebar^2}{\Gep \kp^2 +\Gpp \ke^2 + k \ke \kp/\lep}, 
\label{eq:G_eff_steady_state_general}
\end{equation}
which generalizes the expression derived in~\cite{majumdar2004} to the case where 
$\Gep \neq 0$. In particular, one can see that the effect of electron-phonon coupling may be to increase or decrease the phonon-phonon resistance $\Gpp$ depending on the sign and amplitude of the last term of Eq.~\eqref{eq:G_eff_steady_state_general}.

At this point, it is useful to consider the limit $\kpbar \ll \kebar$ 
and $\kp \Gep \ll \ke \Gpp$. Under these conditions, the 
expression of the effective thermal conductance simplifies:
\begin{equation}
\Geff \simeq \Gpp + \Gep - \Gpp^2 \lep/\ke, 
\label{eq:G_eff_steady_state_appx}
\end{equation}
where we have assumed $\Gpp \ll \sqrt{\kp G}$, 
an approximation valid when 
$\Gpp < 10^8$ \bcunit, 
$\kp> 10$  W$\,$m$^{-1}\,$ K$^{-1}$ and $G> 10^{17}\,$W/m$^{3}$/K.
Under this approximation, the last term of Eq.~\eqref{eq:G_eff_steady_state_appx} 
$\Gpp^2 \lep/\ke=\Gpp^2 \sqrt{\kp/G}$ is small compared to $\Gpp$ and the effect of electron-phonon processes is to increase the total thermal conductance with respect to $\Gpp$:~$\Geff \simeq \Gpp + \Gep$.


\section*{Appendix B:~numerical solution of the two-temperature model}
\label{sec:appendix_B}
\subsection*{Model and parameters}

\begin{figure}[t!]
\includegraphics[width=8cm]{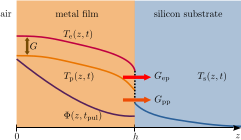}
\caption{
\corf{
Schematic illustration of the two-temperature model (TTM). The metal film has thickness $h$ and the silicon substrate is infinite.  
$\Te$, $\Tp$ and $\Ts$ denote the temperature of metal electron, metal phonon and silicon phonon, respectively. $G$ is the electron-phonon coupling constant, 
while $\bc$ and $\bce$ are the two interfacial conductances. 
$\Phi(z,t)$ is the local heat source for electron. 
}
}
\label{fig:schematic2Tmodel}
\end{figure}

\begin{figure*}[htbp]
\includegraphics[width=16cm]{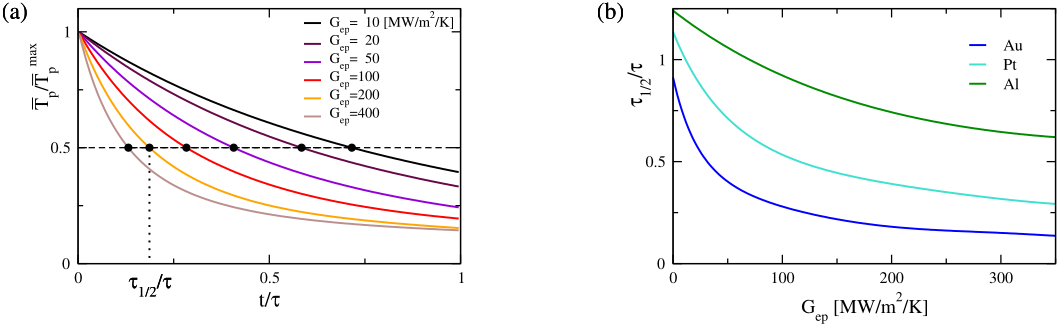}
\caption{
Relaxation time in the two-temperature model.
(a)~Average phonon temperature $\Tpav(t)$ as a function of time in the Au system, 
for various values of electron-phonon conductance~$\bce$. 
Temperature is rescaled by its maximum value $\Tpavmax$ and 
time is rescaled by the characteristic time $\tau=6.4\,$ns defined in Eq.~\eqref{eq:tau}. 
The decay time~$\tauh$ is indicated by dots. 
(b)~Decay time~$\tauh$ computed as a function of electron-phonon conductance~$\bce$, 
in the Au, Pt and Al systems. 
The characteristic time is  $\tau=6.4$, $4.4$ and $1.95\,$ns respectively. 
}
\label{fig:2Tmodel}
\end{figure*}

We consider a two-temperature model, treated in a one-dimensional geometry, 
\corf{as illustrated in Fig.~\ref{fig:schematic2Tmodel}}. 
Here and contrary to the situation considered in  Appendix A, we do not assume steady state conditions. Specifically, we model the thermal relaxation in a metal thin film on a substrate, following the excitation by a laser pulse.  
The evolution equations, which implies diffusion and exchange, read as
\begin{subequations}
\begin{align}
\ce(\Te) \, \partial_t \Te &=  \ke \, \partial_{zz}^2 \Te \, - \, G(\Te - \Tp) + \Phi(z,t),  \\
\Cp      \, \partial_t \Tp &=  \kp \, \partial_{zz}^2 \Tp \, + \, G(\Te - \Tp),               \\
\cs      \, \partial_t \Ts &=  \ks \, \partial_{zz}^2 \Ts.                      
\end{align}
\end{subequations}
Here, 
$T \eqdef \Tab - \To$ is the difference between the absolute temperature~$\Tab$
and a reference temperature~$\To$. 
The subscripts $m=\el$, $\ph$, and $\phs$ refer  respectively 
to the metal electrons, the metal phonons and the substrate phonons. 
$c_m$ and $k_m$ are the heat capacities and thermal conductivities, 
and $G$ is the electron-phonon coupling constant. 
All material coefficients are assumed constant, 
except for the electron heat capacity which is temperature dependent. 
Specifically, we employ the relationship  
$\ce(\Te) = \gamma \Teab = \gamma (\To + \Te)$. 
Finally, $\Phi(x,t)$ is the heat source for electrons induced by laser illumination. 

The boundary conditions for the metallic domain are
\begin{subequations}
\begin{align}
z=0,  \quad  -\ke \, \partial_z \Te  &= -\kp \, \partial_z \Tp =0,                                \label{eq:bc0}     \\ 
z=h,  \quad  -\ks \, \partial_z \Ts  &= \bce (\Te-\Ts) \nonumber \\ 
 & + \bc (\Tp-\Ts),                          \label{eq:bch1}    \\
                                    &= -\ke \,  \partial_z \Te \: - \: \kp  \,  \partial_z \Tp.  \label{eq:bch2}  
\end{align}
\end{subequations}
Equation~\eqref{eq:bc0} assumes an air\itf metal interface that is isolating,  
i.e. forbidding any energy flux. 
Equations~\eqref{eq:bch1}-\eqref{eq:bch2} indicate that 
heat transfer across the metal\itf substrate interface occurs in two ways: 
the phonon-phonon and electron-phonon channels, characterized by conductance $\bc$ and $\bce$ respectively. 
Before the laser pulse, the entire system is at ambient temperature~$\To$, 
thus giving the initial conditions: 
\begin{align}
\mbox{for\ } t=0,  \quad \Tein =\Tpin=\Tsin=0. 
\end{align}
As regards the heat source for electrons, it is given by 
\begin{align}
 \Phi(z,t) = \am \, \frac{h  e^{-z/\delta}}{\delta (1 -e^{-h/\delta})} \, \gaus(t-\tpul, \taupul), 
\end{align}
where $\am$ is the amplitude and $\delta$ the penetration depth 
characterizing the exponential decay.  
$\gaus(u,\varsigma)$ is a normalized centered Gaussian with variable~$u$ and variance~$\varsigma$, 
while $\tpul$  and $\taupul$ give respectively the time and duration of the pulse. 
The form chosen for $\Phi(z,t)$ ensures that 
the total energy deposited by the pulse in the electron
is specified by the amplitude~$\am$

As regards numerical parameters, 
the material properties are gathered in Tab.~\ref{Table_parameters}. 
The metal thickness is $h=100\,$nm and the penetration depth is $\delta=20\,$nm~\footnote{
The substrate thickness is $\hs = 50\,h$, 
which is large enough to model an infinite substrate 
on the time scale of order~$\tau$.}.    
The pulse is characterized by a duration $\taupul=1\,$ps,   
a maximum at $\tpul=2\taupul$ 
and an amplitude~$\am$  chosen so that the electron maximal temperature 
is around $3\To$, with $\To=293\,$K  the ambient temperature. 
For later use, we introduce 
\begin{align}
 \tau = \frac{\Cp h}{\bc},  \label{eq:tau}
\end{align}
which is the typical time for temperature relaxation in the system~\footnote{
Assuming a uniform metal phonon temperature $T(t)$, no electron-phonon conductance ($\bce=0$) 
and a substrate temperature fixed to $\To$, we have $\partial_t T(t)= - \tau^{-1} T(t)$ and 
$T(t) \sim \exp(-t/\tau)$.}. 
For Au, Pt and Al systems, one gets $\tau=6.4$, $4.4$ and $1.95\,$ns respectively. 
Because those values are different,   
it is convenient to rescale the decay time by $\tau$ 
when comparing the three systems. 

\subsection*{Using decay time $\tauh$ to infer $\bce$}

Using numerical resolution of the two-temperature model, 
we characterize the relaxation dynamics of phonon temperature. 
Let us introduce $\Tpav(t)$ as the spatial average temperature of metal phonon 
and $\Tpavmax$ as its maximal value. 
We choose to characterize the relaxation dynamics 
by the decay time~$\tauh$ at which phonon temperature has decreased to half its maximal value, that is 
\begin{align}
 \Tpav(t=\tauh) = \frac{\Tpavmax}{2}. 
\end{align}
Fixing $\bc$ to its ab-initio value determined in Sec.~\ref{sec:comparison} above, 
we show in Fig.~\ref{fig:2Tmodel}a 
the decay of phonon temperature  for various values of electron-phonon conductance~$\bce$. 
As the energy transfer in this channel becomes more efficient, 
the relaxation becomes faster. 
This dependence is quantified by the decay time $\tauh$, 
which is shown in Fig.~\ref{fig:2Tmodel}b as a function of~$\bce$ 
for all three systems. 

\begin{figure}[htbp]
\includegraphics[width=7cm]{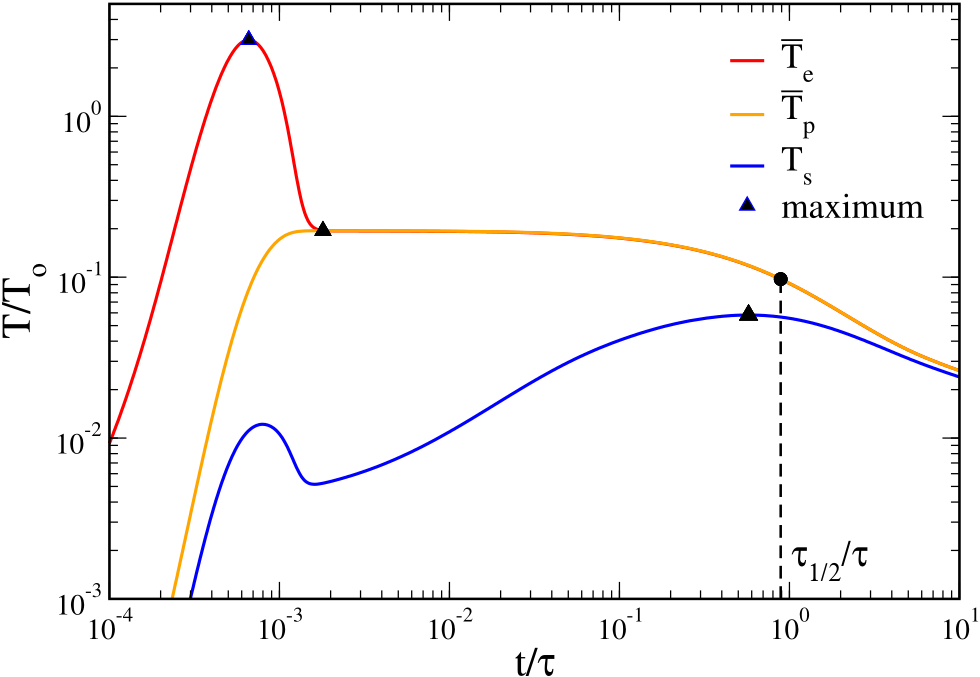}
\caption{
Temperature dynamics in the two-temperature model of Al/Si system. 
Numerical solution for the average temperature $\Teav$  and $\Tpav$ of metal electron and phonon, 
and the substrate temperature at interface $\Ts(t,z=h)$. 
Temperature is rescaled by $\To$ and time by $\tau=1.95\,$ns defined in Eq.~\eqref{eq:tau}. 
The phonon conductance is  ~$\bce=210\,$\bcunit{}   
and the electron conductance is the inferred value~$\bce=107\,$\bcunit.  
Points indicate the maxima (triangles) and the decay time~$\tauh$ (circle). 
}
\label{fig:ex}
\end{figure}

We now use the two-temperature model and experimental data 
to infer the value of electron-phonon conductance. 
Let us assume an experiment where~$\tauh$ was measured 
and interpreted with a two-temperature framework but without electron-phonon conductance ($\bce=0$), 
thus giving an apparent phonon-phonon conductance $\bcexp$. 
Given a theoretical prediction for the phonon-phonon conductance, 
can we infer from experimental data a value for~$\bce$? 
We propose to do so by equalizing the decay time~$\tauh$ 
in the two situations and seek~$\bce$ that satisfies 
\begin{align}
 \tauh (\bcth,\bce) = \tauh (\bcexp,\bce=0). 
\end{align}
The left-hand side is shown in Fig.~\ref{fig:2Tmodel}b 
and the right-hand side (not shown) was computed, 
taking for $\bcexp$ the values from experimental measurements (Fig.~\ref{fig:ne_all}). 
Specifically, we take  $\bcexp =43$  and 350  \bcunit{} for Au/Si and Al/Si interfaces respectively, and $\bcexp$  in the interval [80-150]  \bcunit{} for the Pt/Si interface.  
The values of electron-phonon conductance deduced through this procedure are provided in the main text as Eqs.~\eqref{eq:GepfromTTMa}-\eqref{eq:GepfromTTMc}. 
Note that our estimates for the conductance are not tied 
to the specific choice of $\tauh$. 
Defining $\tau_\chi$ as the time needed for the temperature 
to be reduced by a factor $\chi<1$, we find that 
varying $\chi$ in the range $[0.4-0.8]$ gives  very similar results.  
Specifically, for  Au/Si  and Pt/Si, the deduced value $\Gep$ varies only  by a few \bcunit, while for Al/Si, it varies from 104 to $110\,$\bcunit,  close to the value 107 \bcunit{} reported for $\chi=1/2$.

%
For illustration purpose and taking the Al/Si system as an example,  
we show in Fig.~\ref{fig:ex} the temperature dynamics of all components. 
Several remarks are in order. 
The maxima in electron  and phonon temperature, 
as well as the equilibration time where phonon and electron reach similar values,  
are all much shorter than the characteristic time~$\tau$. 
There is thus a clear separation of time scale 
between metal heating  and  cooling, 
as visible in the large plateau maintained over approximated two decades in time.  
Besides, 
the decay time~$\tauh$ is on the order of~$\tau$ as expected~\footnote{
This expectation is warranted as long as $\bce$ is small or comparable to $\bc$, 
but would not hold if $\bce \gg \bc$.}. 
Finally, the substrate temperature at the interface, $\Ts(t,z=h)$, 
also reaches a maximum at a time of order~$\tau$.  


\bibliography{apssamp}

\end{document}